\pdfoutput=1

\documentclass[authorversion, nonacm]{acmart}

\AtBeginDocument{%
  \providecommand\BibTeX{{%
    \normalfont B\kern-0.5em{\scshape i\kern-0.25em b}\kern-0.8em\TeX}}}

\usepackage{graphics} 
\usepackage{caption}
\usepackage{dirtytalk}
\usepackage{subfigure}
\usepackage{amsmath}
\usepackage{wrapfig}

\newcommand{\TM}{\textsuperscript{TM}}

\begin{document}


\title [Recognizing Complex Gestures on Minimalistic Knitted Sensors] {Recognizing Complex Gestures on Minimalistic Knitted Sensors: Toward Real--World Interactive Systems}

\author{Denisa Qori McDonald}
\email{denisaqori@gmail.com}
\affiliation{%
    \institution{College of Computing and Informatics, Drexel University}
    \streetaddress{3675 Market St.}
    \city{Philadelphia}
    \state{Pennsylvania}
    \country{USA}
}

\author{Richard Vallett}
\email{rjvallett@drexel.edu}
\affiliation{%
    \institution{Center for Functional Fabrics, Drexel University}
    \streetaddress{3101 Market St.}
    \city{Philadelphia}
    \state{Pennsylvania}
    \country{USA}
}

\author{Lev Saunders}
\email{levsau00@gmail.com }
\affiliation{%
  \institution{College of Computing and Informatics, Drexel University}
    \streetaddress{3675 Market St.}
    \city{Philadelphia}
    \state{Pennsylvania}
    \country{USA}
}

\author{Genevi\`{e}ve Dion}
\affiliation{%
    \institution{Center for Functional Fabrics, Drexel University}
    \streetaddress{3101 Market St.}
    \city{Philadelphia}
    \state{Pennsylvania}
    \country{USA}
}

\author{Ali Shokoufandeh}
\affiliation{%
    \institution{College of Computing and Informatics, Drexel University}
    \streetaddress{3675 Market St.}
    \city{Philadelphia}
    \state{Pennsylvania}
    \country{USA}
}

\renewcommand{\shortauthors}{McDonald, et al.}

\begin{abstract}
Developments in touch-sensitive textiles have enabled many novel interactive techniques and applications. Our digitally-knitted capacitive active sensors can be manufactured at scale with little human intervention. Their sensitive areas are created from a single conductive yarn, and they require only few connections to external hardware. This technique increases their robustness and usability, while shifting the complexity of enabling interactivity from the hardware to computational models. This work advances the capabilities of such sensors by creating the foundation for an interactive gesture recognition system. It uses a novel sensor design, and a neural network-based recognition model to classify 12 relatively complex, single touch-point gesture classes with 89.8\% accuracy, unfolding many possibilities for future applications. We also demonstrate the system's applicability and robustness to real-world conditions through its performance while being worn and the impact of washing and drying on the sensor's resistance. 

\end{abstract}

\ccsdesc[500]{Human-centered computing~Gestural input}
\ccsdesc[300]{Human-centered computing~Interaction devices}
\ccsdesc[500]{Computing methodologies~Neural networks}
\ccsdesc[300]{Computing methodologies~Cross-validation}
\ccsdesc[300]{Computing methodologies~Supervised learning by classification}
\ccsdesc[300]{Human-centered computing~Ubiquitous computing}
\ccsdesc[300]{Human-centered computing~User studies}

\keywords{gesture detection, knitted sensors, wearable sensor, CNN networks, LSTM networks, neural networks, evaluation study, interactive system, real-time system, real-world conditions}


\begin{teaserfigure}
    \subfigure[]
    {\includegraphics[height=1.25in]{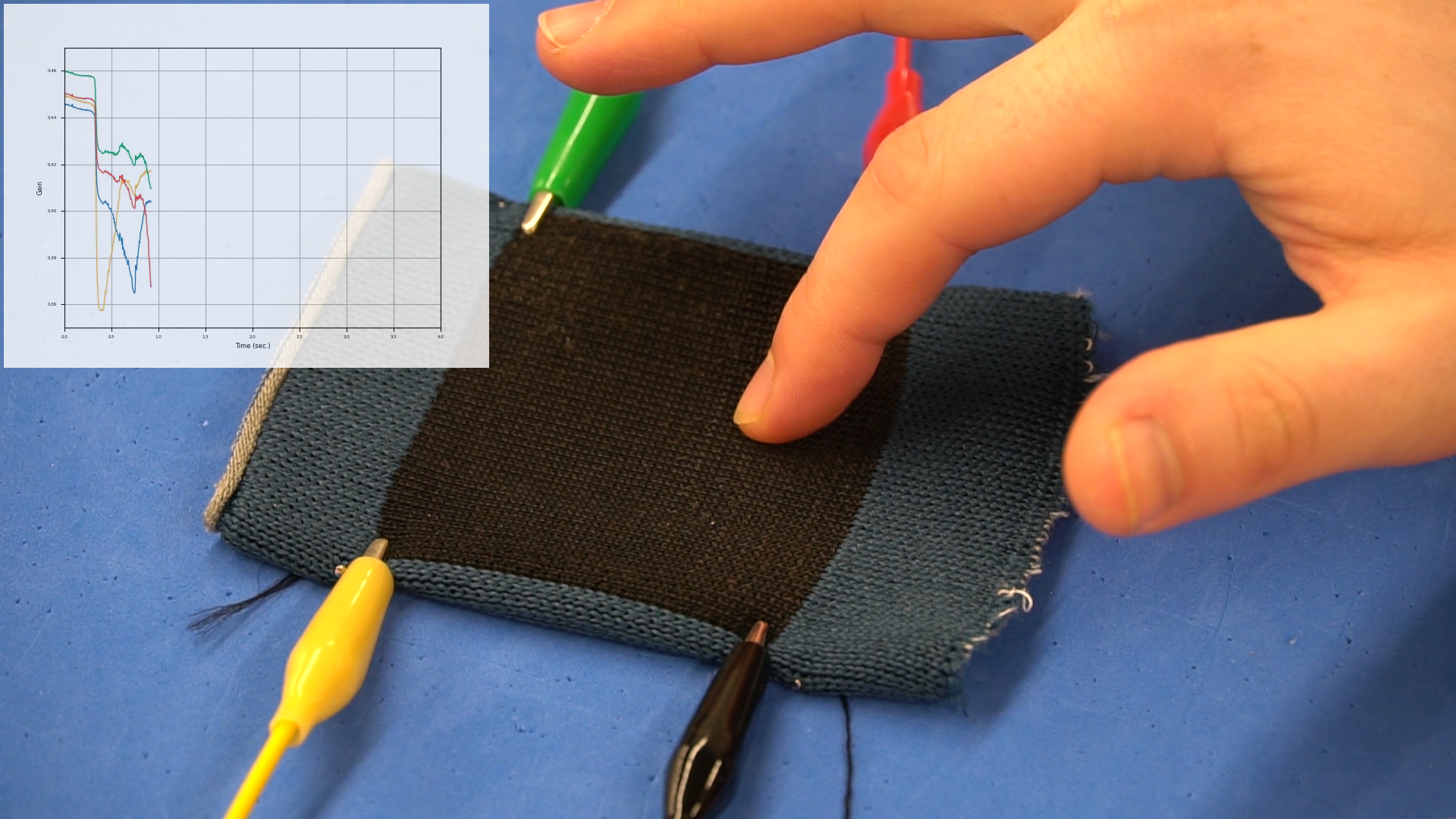}\label{fig:sensor_electrodes}}
    \hfill
    \subfigure[]
    {\includegraphics[height=1.25in]{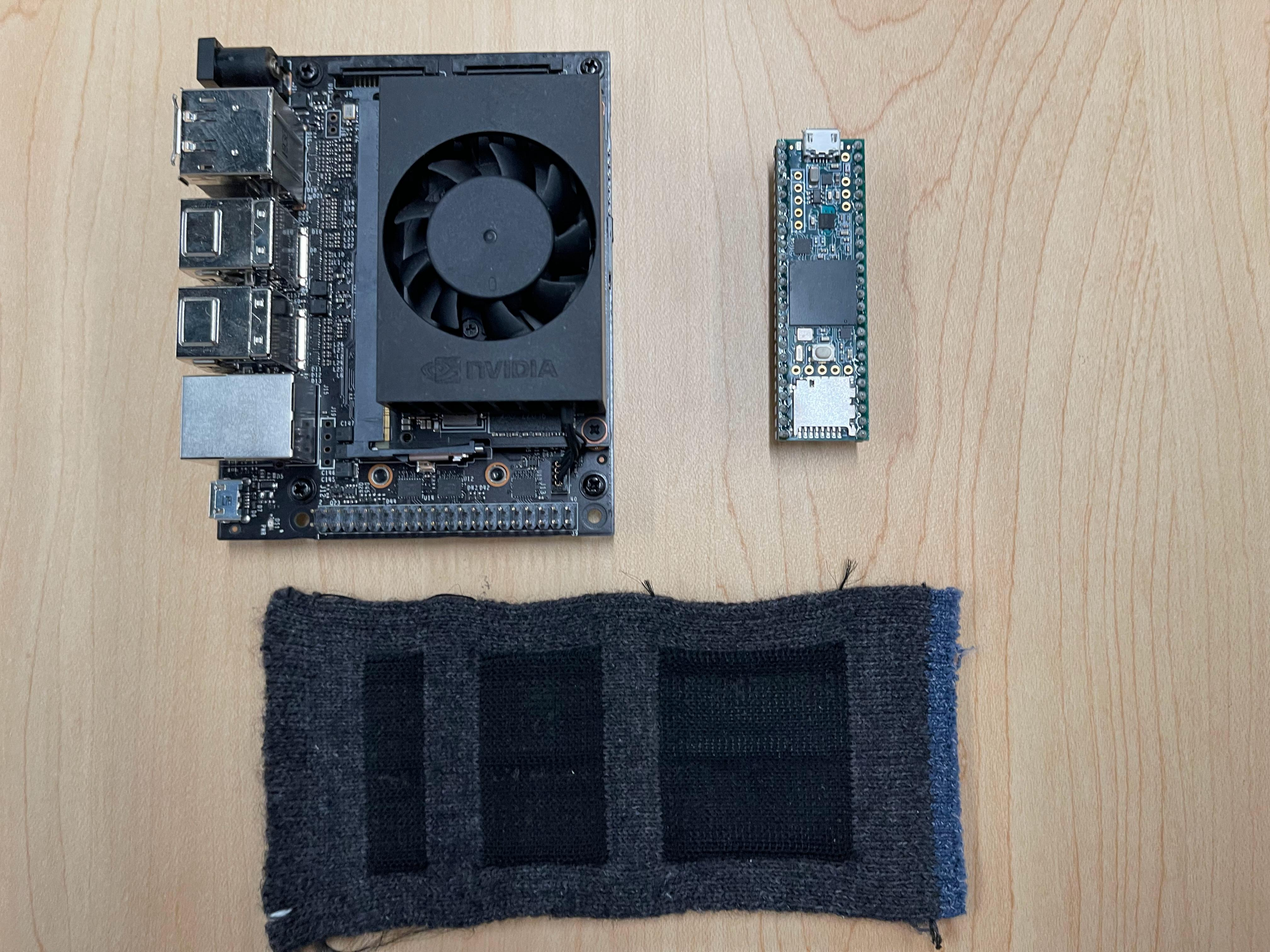}\label{fig:sensor_stretch}}
    \hfill
    \subfigure[]
    {\includegraphics[height=1.25in]{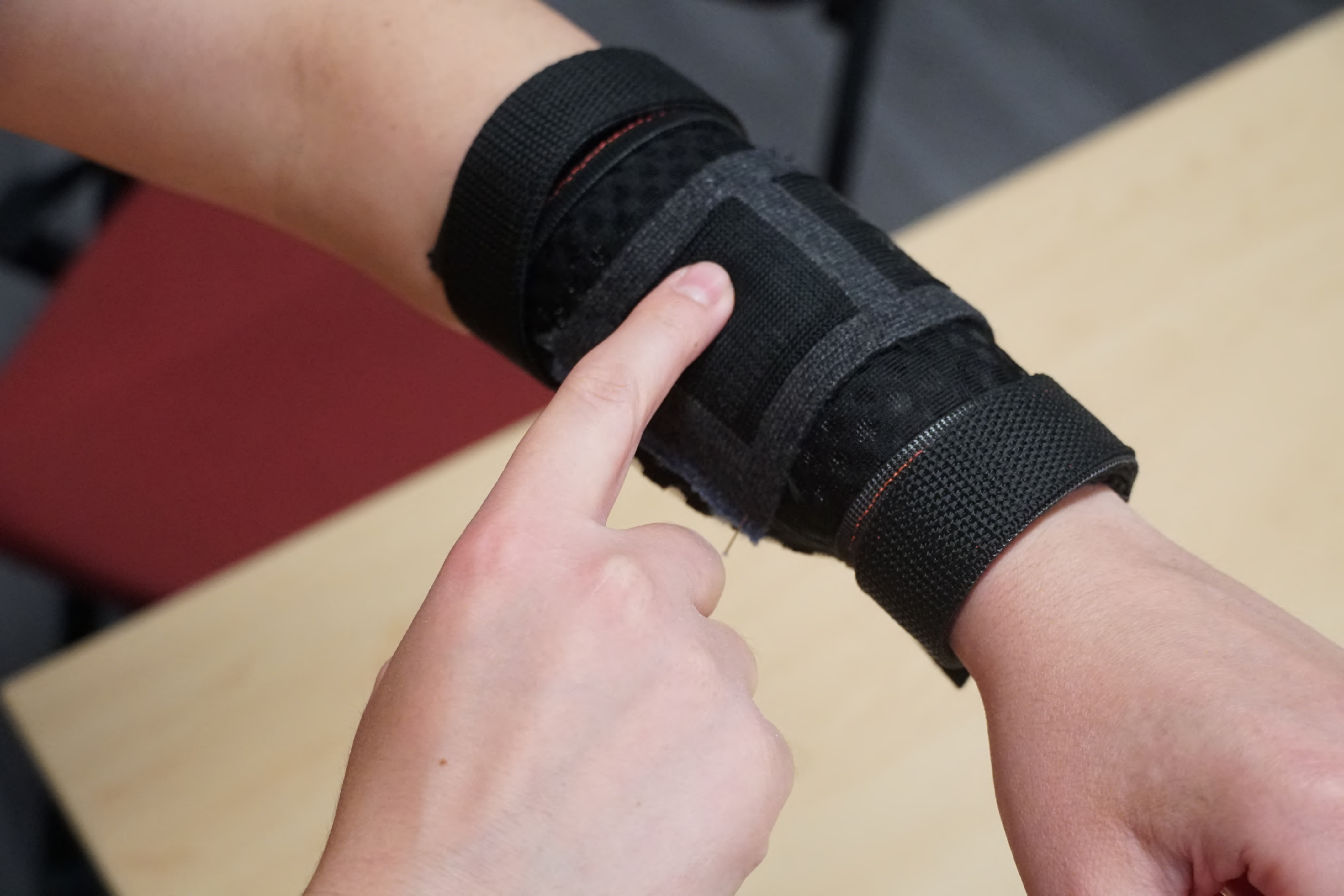}\label{fig:wearable_sensor}} 
    \caption{Knitted sensors for gesture recognition, constructed with carbon-coated nylon yarn (darker rectangular regions) and polyester yarn. \emph{(a)}: A knitted sensor with four electrodes attached capturing gesture information. \emph{(b)}: Components to construct future real-time gesture recognition systems: an NVIDIA® Jetson Xavier™ NX Developer Kit hosting a trained model, a micro-controller for signal generation and processing, and the knitted sensor for gesture input. The knitted component is a similarly designed and constructed sensor as in \emph{(a)}, but with three smaller conductive areas. \emph{(c)}: The same knitted sensor shown in \emph{(b)} being worn.} 
    \Description{Three photos labelled (a), (b) and (c) show images of a knitted sensor under different circumstances. Image (a) shows a sensor attached to four electrodes and a user touching it with their finger. In (b) a similarly constructed knitted sensor to the one in (a) is shown alongside a micro-controller and an NVIDIA system-on-module computer. In (c) a user is wearing the same knitted sensor shown in (b) on their forearm, an touching it with one finger.}
\end{teaserfigure}

%

\maketitle

\newcommand{\DeltaRBDryOneAB}{$-6.12\%$}
\newcommand{\DeltaRBDryOneAC}{$28.09\%$}
\newcommand{\DeltaRBDryOneAD}{$8.52\%$}
\newcommand{\DeltaRBDryOneBC}{$12.32\%$}
\newcommand{\DeltaRBDryOneBD}{$8.93\%$}
\newcommand{\DeltaRBDryOneCD}{$1.77\%$}
\newcommand{\DeltaRBDryTwoAB}{$7.92\%$}
\newcommand{\DeltaRBDryTwoAC}{$31.48\%$}
\newcommand{\DeltaRBDryTwoAD}{$29.84\%$}
\newcommand{\DeltaRBDryTwoBC}{$31.26\%$}
\newcommand{\DeltaRBDryTwoBD}{$20.52\%$}
\newcommand{\DeltaRBDryTwoCD}{$15.44\%$}
\newcommand{\DeltaRBDryThreeAB}{$0.77\%$}
\newcommand{\DeltaRBDryThreeAC}{$8.82\%$}
\newcommand{\DeltaRBDryThreeAD}{$14.08\%$}
\newcommand{\DeltaRBDryThreeBC}{$18.42\%$}
\newcommand{\DeltaRBDryThreeBD}{$17.98\%$}
\newcommand{\DeltaRBDryThreeCD}{$0.55\%$}
\newcommand{\DeltaRBDryFourAB}{$7.05\%$}
\newcommand{\DeltaRBDryFourAC}{$15.36\%$}
\newcommand{\DeltaRBDryFourAD}{$17.09\%$}
\newcommand{\DeltaRBDryFourBC}{$20.18\%$}
\newcommand{\DeltaRBDryFourBD}{$13.20\%$}
\newcommand{\DeltaRBDryFourCD}{$-2.65\%$}
\newcommand{\DeltaRBDryFiveAB}{$-2.10\%$}
\newcommand{\DeltaRBDryFiveAC}{$11.07\%$}
\newcommand{\DeltaRBDryFiveAD}{$7.42\%$}
\newcommand{\DeltaRBDryFiveBC}{$11.03\%$}
\newcommand{\DeltaRBDryFiveBD}{$-3.80\%$}
\newcommand{\DeltaRBDryFiveCD}{$-2.21\%$}
\section{Introduction}
Innovation based on fabric sensors has the potential to enable many interactive applications. Through machine learning, this work advances gesture recognition in capacitive knitted sensors, designed to rely on one conductive yarn and few external connections. The ultimate goal in developing the underlying technology of these sensors is allowing them to behave like fabric in the real world~\cite{gemperle1998design}, while producing accurate real-time outputs that model complex behavior, upon which it is possible to develop interactive applications. 

Large-scale manufacturing outside a lab is important for real-world relevance of this technology, and has been addressed by many~\cite{Poupyrev2016a,chen2020design,Wicaksono2020KnittedKeyboard,Vallett2016a,Vallett2019a}. In order to create a sensor less susceptible to malfunctioning, needing few fabric-to-wire connections, and able to be manufactured at scale through an easily repeatable process, we use digital weft knitting, in a process similar to other work~\cite{Vallett2016a, Vallett2019a, mcdonald2020knitted}. A conductive, carbon-coated nylon yarn is routed together with other yarns in the fabric construction process according to a pre-programmed design pattern. There are four external connection points to this fabric component.

The limited number of external connections in these sensors makes manufacturing easier, while improving the sensors' robustness and flexibility, which in turn enhance their usability. However, this minimalism also creates the need for more complex computational models to process reduced information output from the system. We rely on a combination of Convolutional Neural Networks (CNN) and Long Short-Term Memory (LSTM) neural network architectures to capture the important aspects of the signal generated from different gestures. CNNs learn local relationships, while LSTMs focus on the sequential aspect of the time series signal data. Prior work~\cite{Vallett2019a, mcdonald2020knitted} has explored accurate touch location identification on a 36-button knitted touchpad, which uses the same design principles and construction process. However, those systems still do not offer gesture recognition capabilities.

This work builds the foundations for creating an \emph{interactive gesture recognition system}, which would enable many applications. We introduce a \emph{sensor pattern} designed, through programmatically routing the carbon-fiber yarn, for gesture input, containing one solid sensing area and 4 electrodes to connect to external hardware; a \emph{supervised CNN-LSTM neural network model} to classify 12 relatively complex gestures performed on a knitted sensor, which are a subset of the characters of English language; \emph{results from three user studies} for training, evaluation of the model under normal lab conditions, and evaluation of the model while the sensor is being worn. Additionally, to get closer to real-world use, we present \emph{results from an experiment} investigating the effect of washing and drying on the sensor's resistance, an important property related to its circuit design and ultimately trained machine learning model.  

In the section below, we describe related work, followed by introducing our gesture recognition system composed of the knitted component, signal measurement circuit, signal processing pipeline, the gesture recognition model, and the deployment hardware. Our user studies, in the successive section, serve to collect data for training the model, and to evaluate its stability with new data. Next, we explore the applicability of this work in the real world, by furthering its technical evaluations and discussing the necessary processes and components to build applications relying on the gesture recognition capabilities of such a system. Following these explorations, the remaining sections provide more context, limitations, and future directions.

\section{Related Work}
This section starts by providing some context regarding fabric touch sensors produced using different fabric construction or embellishment techniques, as well as different sensing methods, focusing on capacitive sensing upon which this system is built. Then, the discussion transitions to work that has focused on gesture recognition on fabric sensors, which have been produced through various techniques. Subsequently, a discussion of the neural network architectures upon which our model relies, provides an overview of their working and the reasons for our choice.

\subsection{Smart Textile Construction Processes}

\begin{figure*}[h]
    \centering
    \subfigure[]
    {\includegraphics[width=0.18\textwidth]{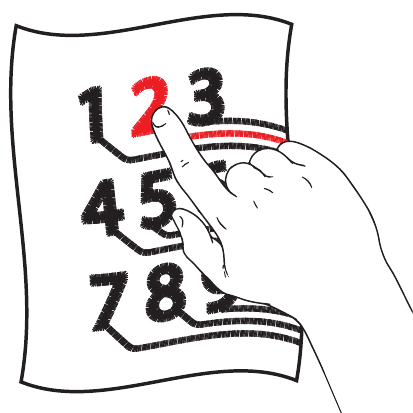}\label{fig:Embroidered_Sensor}}
    \hfill
    \subfigure[]
    {\includegraphics[width=0.18\textwidth]{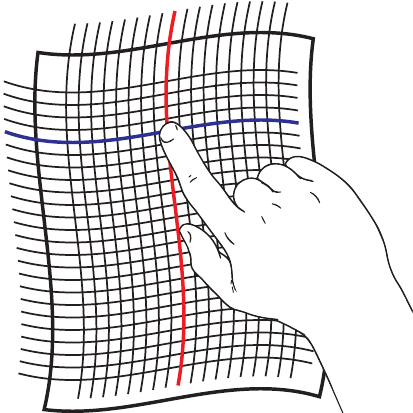}\label{fig:Capacitive_Sensing_Matrix}}
    \hfill
    \subfigure[]
    {\includegraphics[width=0.18\textwidth]{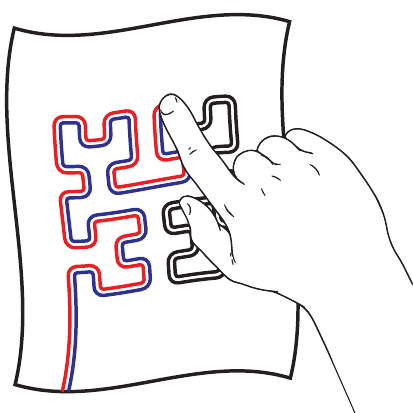}\label{fig:Reflectometry}}
    \hfill
    \subfigure[]
    {\includegraphics[width=0.18\textwidth]{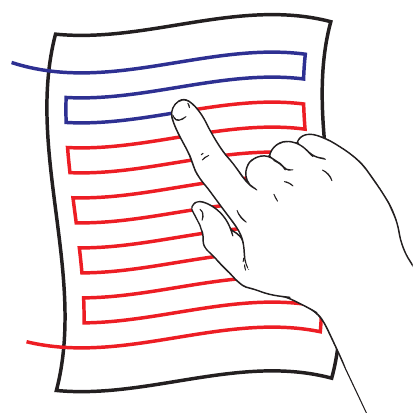}\label{fig:Knitted_Sensor}}
    \hfill
    \caption{Illustrations of sensing techniques and electrode layouts used in textile touch sensors. \emph{(a)}: Embroidered capacitive buttons. \emph{(b)}: Woven or embroidered capacitive sensing matrix. \emph{(c)}: \emph{Reflectometry} measured using two parallel conductors. \emph{(d)}: Single weft knitted conductor measured using \emph{differential capacitive sensing}.}
    \label{fig:Designs}
    
    \Description{The four photos, (a), (b), (c), and (d) show different sensing strategies used for fabric sensors. On each figure, there is an illustrated hand touching on an area of the illustrated fabric sensor. In (a), there are nine buttons shown, from 1-9, with all of them colored black, except for number 2, which is red, illustrating the button being touched. Figure (b) shows a matrix of vertical and horizontal lines. There is a vertical blue line and red horizontal line intersecting on the point of touch. In (c), there are two parallel lines forming a shape together. One of the lines is red up to the point of touch on the fabric, and black after it; the other line is completely black. Figure (d) shows a serpentine line, where the top part of the line, which is before the point of touch, is blue, while the bottom one, after the point of touch is red.}
\end{figure*}

Textile assembly or embellishing methods, such as embroidery, weaving, and knitting have been used to create touch-sensitive textile devices using conductive yarn. \emph{Embroidery} is a textile embellishing method that inserts thread into a fabric substrate to create arbitrary patterns (Figure~\ref{fig:Embroidered_Sensor}), and has been used to create several interactive textile devices~\cite{Post2000a,Gilliland2010a,Hamdan2016a,Hamdan2018a,aigner2020embroidered,mlakar2020design}. Embroidered threads are not part of the original textile, therefore they can be overlaid to form a variety of sensing shapes on any textile, if conductive yarn in used. Large-area distributed touch sensing can also be achieved through integrating conductive yarns into the textile structure during formation, eliminating post processing. {\em Weaving} and {\em knitting} are textile manufacturing processes that can combine conductive and non-conductive yarns to form fabric circuits.

In {\em weaving}, a textile is formed by interlacing perpendicular yarns~(Figure~\ref{fig:Capacitive_Sensing_Matrix}). It is a widely-used method to produce distributed textile touch sensors~\cite{DeRossi2002a,Hasegawa2007a,Takamatsu2011a,Agcayazi2016a,Poupyrev2016a,wu2020zebrasense} due to its similarity to standard capacitive and resistive sensing matrices. Woven textiles use the perpendicular alignment of numerous horizontal and vertical yarns to form a sensing grid and detect touch at yarn intersection points, supporting a high-resolution sensing area. However, connecting a matrix to external sensing electronics requires an equal number of fabric-to-wire connections---a difficult and time-consuming post production process. These connections often exist at millimeter-scales and utilize techniques like soldering, which create brittle joints that reduce the textile's durability, and potentially real-world practical use. 

\emph{Knitting}, the technique used to construct our sensor, produces fabric by employing a continuous yarn or set of yarns to form a series of interlocking loops. Knitting has been recently explored in producing interactive textiles, such as force-sensing knitted structures~\cite{Pointner2020KnitedRESi}, knitted electronic musical controllers~\cite{Wicaksono2020KnittedKeyboard}, and knitted dynamic displays~\cite{devendorf2016don}. Other research has investigated creating 2D sensing structures using a homogeneous~\cite{Vallett2016a,Vallett2019a} or multi-material~\cite{ozbek2018novel} 1D filament. In work more closely related to ours, digital weft-knitting has been explored to construct textiles with only two connection points~\cite{Vallett2016a,Vallett2019a,mcdonald2020knitted}. We select this process since it uses continuous yarns in the production of textiles, which allows the construction of a fabric circuit through an electrically continuous trace. The conductive element in this work is carbon-coated nylon yarn, which becomes intertwined with regular, non-conductive yarns according to a programmatically-defined pattern of textile construction. The circuit design is compatible with the way yarns are routed during the machine knitting process, and does not require post-production assembly, besides attaching wires at the connection points. This process allows for different textile patterns to be produced according to application specifications, as well as similar form factors of different scales.  

\subsection{Fabric-based Capacitive Touch Sensing}
Capacitive touch sensing is a widely utilized method of interacting with electronic devices. Capacitive touch sensors measure the presence or proximity of a conductor, such as human skin, through its electromagnetic properties. These sensors can discern nuanced touch and gesture and often require no mechanical deflection of a sensing medium. This mode of interaction is advantageous for use in touch-sensitive textiles, which need a dynamic circuit structure. Alternative methods of distributed touch detection have been pursued in textiles, such as \emph{inter-yarn contact sensing}~\cite{Inaba1996a, karrer2011pinstripe} and \emph{resistive sensing}~\cite{sundholm2014smart, Parzer2018a, pointner2020knitted, honnet2020polysense, freire2017second}.

Textile-based capacitive touch sensing circuits can be categorized as either \emph{discrete} or \emph{continuous}. Discrete touch sensing relies on numerous electrically isolated conductors distributed across a sensitive surface~\cite{Post2000a,Poupyrev2016a,wu2020capacitivo}. The bulk magnitude of capacitance is measured on each conductor, and touch is localized using knowledge of each conductor’s location. Improving the accuracy and resolution of touch localization relies on increasing the density of conductors within a given area. This is often done at the cost of increasing the circuit’s structural complexity and increasing the number of connections from the circuit to sensing hardware. However, the sensing electronics required to scan the discrete electrodes are simpler than what is required in continuous sensing. Textile manufacturing techniques like weaving (Figure~\ref{fig:Capacitive_Sensing_Matrix}) are often used to create capacitive sensing matrices, while standalone buttons can be achieved using textile embellishment methods like embroidery (Figure~\ref{fig:Embroidered_Sensor}).

\begin{figure*}[ht]
    \centering
    \subfigure[]
    {\includegraphics[width=.49\textwidth]{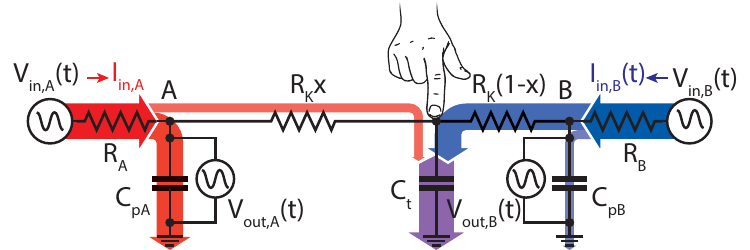}\label{fig:DCS_Circuit_Diagram}}
    \hfill
    \subfigure[]
    {\includegraphics[width=.49\textwidth]{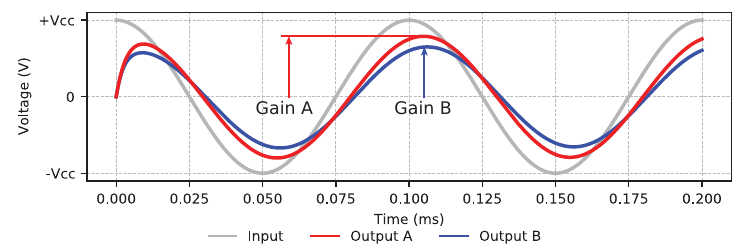}\label{fig:DCS_Waveform_Output}}
    \hfill
    \caption{Illustration of the differential capacitive touch sensing circuit and example waveform output. \emph{(a)}: Circuit diagram visualizing the current differential between the measurement points (A and B) and the location of touch. \emph{(b)}: Example waveform measurements showing the attenuation in voltage (gain A, B) based on the touch location and capacitance magnitude shown in \emph{(a)}. The gains are found as the amplitude ratio of the input and output waveforms using \emph{Bode analysis}.}
    \label{fig:Differential_Capacitive_Sensing}
    
    \Description{The two figures, (a) and (b) show a circuit with two electrode connections, and a graph with three sinusoidal lines. The circuit in figure (a) shows current path on each electrode of the circuit A, in red, and B, in blue. An illustrated hand demonstrates a touch event, upon which the capacitor is shown being discharged in purple. In (b), we see a graph whose X axis shows time (0-200 microseconds), while Y axis shows the voltage values. There are three lines, which represent the input waveform, the output waveform measured at electrode A, and the output waveform measured at electrode B. The gain is measured as the amplitude ratio between the input and measured outputs. Corresponding to the touch location in (a), the waveform from electrode A has a higher value at the peaks compared to electrode B.}
\end{figure*}

In contrast, continuous touch sensing infers the location of touch across a single electrically continuous conductor. Continuous touch sensing exploits the properties of the conductor to measure changes in capacitance along its length, which are localized using knowledge of the conductor’s overall path. The geometry of the conductor may facilitate localization, such as creating textile-based transmission lines for use with Time- and Frequency-domain Reflectometry (TDR and FDR)~\cite{Wimmer2011a,Hughes2014a,Hughes2017a}, as illustrated in Figure~\ref{fig:Reflectometry}, or other more recent methods~\cite{ku2020threadsense}. Other sensing methods like Electric Field Tomography (EFT) measure capacitance along a pattern-less conductive surface using measurements along the periphery~\cite{Zhang2017a}. These methods can localize touch using relatively simple circuits; however, the accuracy and resolution of touch location relies on complex measurement hardware to infer accurate location from recorded data. 

 Previous work has used \emph{differential capacitive sensing}, a continuous capacitive sensing strategy, in conjunction with a weft-knitted circuit composed of a single conductive yarn to localize capacitive touch~\cite{Vallett2016a,Vallett2019a,mcdonald2020knitted}. The circuit design of the sensor in this work is based on that same method. As shown in Figure~\ref{fig:DCS_Circuit_Diagram}, touch location is inferred along a linear conductive pathway via a current differential measured at each endpoint. The measured voltage is visualized in Figure~\ref{fig:DCS_Waveform_Output}, which illustrates the attenuation of voltage based on the location and magnitude of capacitive touch. The waveform gain is determined via \emph{Bode analysis}~\cite{Bode1940a}, which determines the amplitude ratio of the input and output waveforms using Fast Fourier Transforms. Initial applications focused on creating user interfaces capable of measuring simple tap location. Our work extends the structure and capabilities of the sensing circuit by using a planar conductive area to detect continuous gestures.  

\subsection{Machine Learning Applications for Fabric-based Touch Sensing}
While textile touch sensors can provide basic touch location, machine learning algorithms have been used to enable higher-level, accurate input recognition. \emph{Project Jacquard}~\cite{Poupyrev2016a} introduces a platform to produce interactive woven sensors. In its user evaluation study, gestures such as \emph{swipe left}, \emph{swipe right}, and \emph {hold} were recognized under three different condition: \emph{sitting}, \emph{standing}, and \emph{walking}. The work from Hughes et al.~\cite{Hughes2014a,Hughes2017a} describes the construction of an RF-based e-Textile and its ability to distinguish between gestures of \emph{tap}, \emph{up swipe}, and \emph{down swipe} based on a CNN model. \emph{SmartSleeve}~\cite{Parzer2017a}, a deformable, pressure-sensing textile sensor recognizes in real time both surface and deformation gestures. It relies on an algorithms which, after converting the data in image form, uses an SVM-trained model together with heuristics to detect the gesture type. \emph{GestureSleeve}~\cite{Schneegass2016a} is a touch-sensing textile sleeve, able to detect stroke-based gestures or taps through the \$P algorithm~\cite{vatavu2012gestures}. \emph{Smart-mat}~\cite{sundholm2014smart} uses a kNN classifier on features built using time series statistical information. \emph{RESi}~\cite{Parzer2018a}, a resistive, touch sensing interface uses SVM classification to detect basic gestures such as single, double and triple taps, as well as swipes in the four directions. More recent work~\cite{olwal2020textile} furthers cord-based interfaces by enabling continuous control as well as recognizing casual discrete gestures.

Our work focuses on detecting gestures that are relatively complex compared to most of these examples. Additionally, all the sensors described above, mainly focused on gesture recognition, differ from our work in construction and sensing strategy. In many of these cases and other similar sensors, much of the complexity lies on the hardware design. In this work however, the construction is kept simple and streamlined, while shifting the burden of complexity to computational models, similarly to McDonald et al.~\cite{mcdonald2020knitted}. In that work, an LSTM neural network with features constructed on statistical information from voltage gain values was used to extract location of touch from a single-yarn and two-connection capacitive knitted sensor. Extending the functionality of that work, we introduce a capacitive knitted sensor, which is also composed of a single yarn, but uses four external connections to recognize gestures. Below, we describe the machine learning concepts upon which our model is built. 

\subsection{CNNs and LSTMs}
To recognize gestures performed on the knitted sensor, we use a neural network architecture which combines Convolutional Neural Networks (CNN)~\cite{fukushima1980neocognitron,lecun1998gradient} with Long Short-Term Memory networks (LSTM)~\cite{gers1999learning}, a stable version of Recurrent Neural Networks (RNNs)~\cite{rumelhart1988learning}. The time-series data, acquired through the four connections in the sensor we design, needs to be processed to map to particular gestures. Since this is time series data, the classification algorithm should take advantage of the continuity of sampled frequencies over time, in order to properly model the necessary connections. Moreover, due to variability in the data, such as different users, conditions, finger placement and more, the same gestures should be represented in terms of essential patterns within them. To those ends, we utilize CNNs and LSTMs, which have found application for many classification tasks. 

CNNs are relatively sparse, regularized neural networks, which have translation-invariant characteristics. They aim to capture hierarchical patterns in data, composing higher-level features from lower-level ones. CNNs have been successfully used, among other areas, in image and video recognition tasks, natural language processing, and recommender systems. They capture the spatial relationships between samples. RNNs allow previous outputs to be utilized as inputs, in addition to hidden states. This flexibility makes them ideal for classification of inputs of any length, while maintaining limits on model size, taking into account historical information, and sharing weights across time. A known drawback of RNN models is their difficulty in capturing long term dependencies due to multiplicative gradients. The error of such models can be exponentially decreasing/increasing with respect to the number of layers. The Gated Recurrent Unit (GRU)~\cite{cho2014learning} and their generalizations, such as Long Short-Term Memory units (LSTM)~\cite{gers1999learning} address the vanishing gradient problem encountered by traditional RNNs. We use LSTMs to model the temporal dependencies in our data. More information is provided regarding our network architecture in the section below, as part of the gesture recognition system.

\section{System Design}
\label{sec:system_design}
In this section, we describe our gesture recognition system, composed of a few main components: the knitted fabric containing conductive and non-conductive yarn, the measurement hardware and circuit, the algorithmic components which produce a trained machine learning model, and ultimately the NVIDIA\textregistered~Jetson Xavier\TM~NX, a powerful embedded system-on-module (SoM) on which the model is deployed. Currently, signal generation, acquisition, and processing occur offline through the use of an arbitrary waveform generator and digital oscilloscope. After the data from multiple users is collected, we train a classification model to distinguish between 12 different gestures. Subsequently, we deploy the model on the embedded CPU and test its ability to recognize a gesture event in real-time. In the future, we plan to perform signal generation, acquisition, and pre-processing on a standalone embedded micro-processor capable of communicating with the embedded CPU, which would allow this system to be fully portable, while providing real-time interaction. 

The selected gestures are the alphanumeric characters: \emph{'3'}, \emph{'5'}, \emph{'I'}, \emph{'J'}, \emph{'L'}, \emph{'M'}, \emph{'O'}, \emph{'S'}, \emph{'V'}, \emph{'W'}, \emph{'Z'}, with the addition of character \emph{'?'}. All of these gestures can be performed in one motion and have a distinct onset and offset. Some are more basic, such as \emph{'I'}, others more complex, such as \emph{'?'}, and some gestures are more similar to each other, such as \emph{'5'} and \emph{'S'}, which could cause incorrect classification. This character subset contains the basic motions upon which other alphanumeric characters can be built. Our choice of several alphanumeric characters serves to strike a balance between prototyping this system, and the complexity it has the potential to offer. Instead of testing its recognition ability with more basic gestures, such as taps and directional swipes, we opted for a more comprehensive set, including gestures with curvatures, to explore the limits of its capabilities. On the other hand, including the whole alphanumeric character set of the English language would increase its complexity even more, requiring more training data. It is worth noting that the focus of this work is not the specific application, but the gesture-enabling technology illustrated through some examples. Ultimately, we would like to build a recognition model capable of detecting numbers and the letters of the alphabet, which could serve as a communication system. Our system is a first step in that direction, but more generally, it unlocks the potential of gesture recognition on knitted sensors with many possible applications. Below we describe its components in more detail.

\begin{figure*}[ht]
    \centering
    \hfill
    \subfigure[]
    {\includegraphics[width=0.4\textwidth]{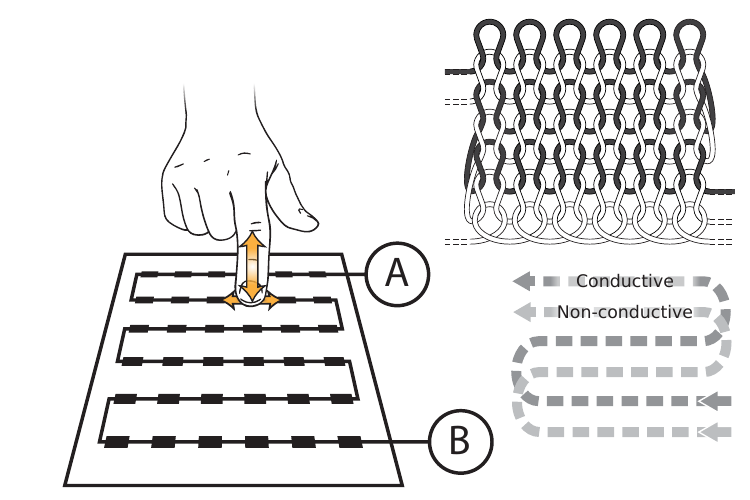}\label{fig:Single-wire_Touchpad_Diagram}}
    \hfill
    \subfigure[]
    {\includegraphics[width=0.4\textwidth]{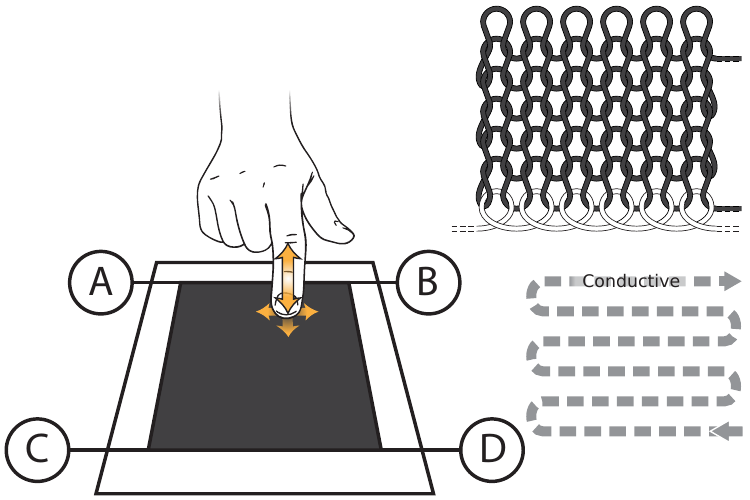}\label{fig:Planar_Touchpad_Diagram}}
    \hfill
    \caption{Diagrams of the knitted touchpad designs. \emph{(a)}: A single-wire touchpad created using conductive and non-conductive yarns that serpentine across the textile surface. The points \emph{A} and \emph{B} connect to external sensing hardware. \emph{(b)}: A touchpad created as a planar conductive area with four connections points, \emph{A, B, C} and \emph{D}.}
    \label{fig:Touchpad_Designs}
    \Description{The two figures, (a) and (b) illustrate how weft-knitted sensors are constructed. In both cases, there is an illustrated hand touching somewhere on the sensor. In (a), the sensing structure is a linear serpentine, and there are two electrode points A and B. An additional view of the construction is provided with intertwined gray and black lines which represent conductive and non-conductive yarns combined to form the fabric. Figure (b) shows a similar sensor; however, the sensing area is a black rectangle in the middle, and there are four connection points A, B, C, and D. The additional view of the construction shows black lines intertwined to form the fabric, representing conductive yarn.}
\end{figure*}

\subsection{Knitted Sensor}
\label{sec:knitted_sensor}

\begin{figure}[ht]
    \centering
    \hfill
    \subfigure[]
    {\includegraphics[width=0.3\textwidth]{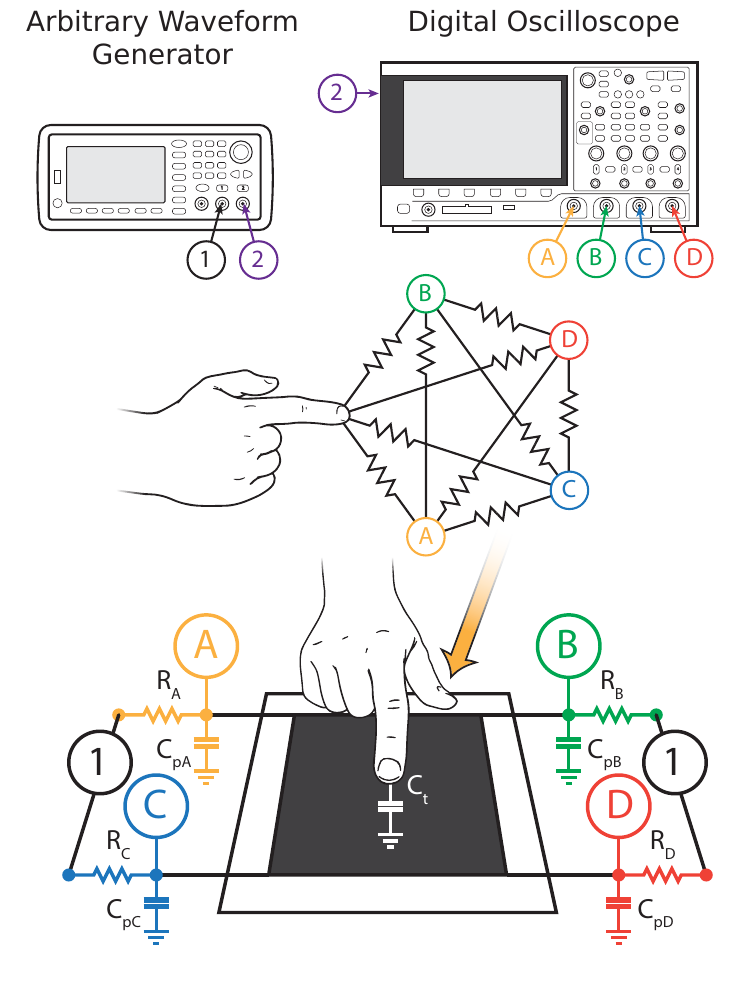}\label{fig:Planar_CTS_Touch_Sensing_Circuit}}
    \hfill
    \subfigure[]
    {\includegraphics[width=0.6\textwidth]{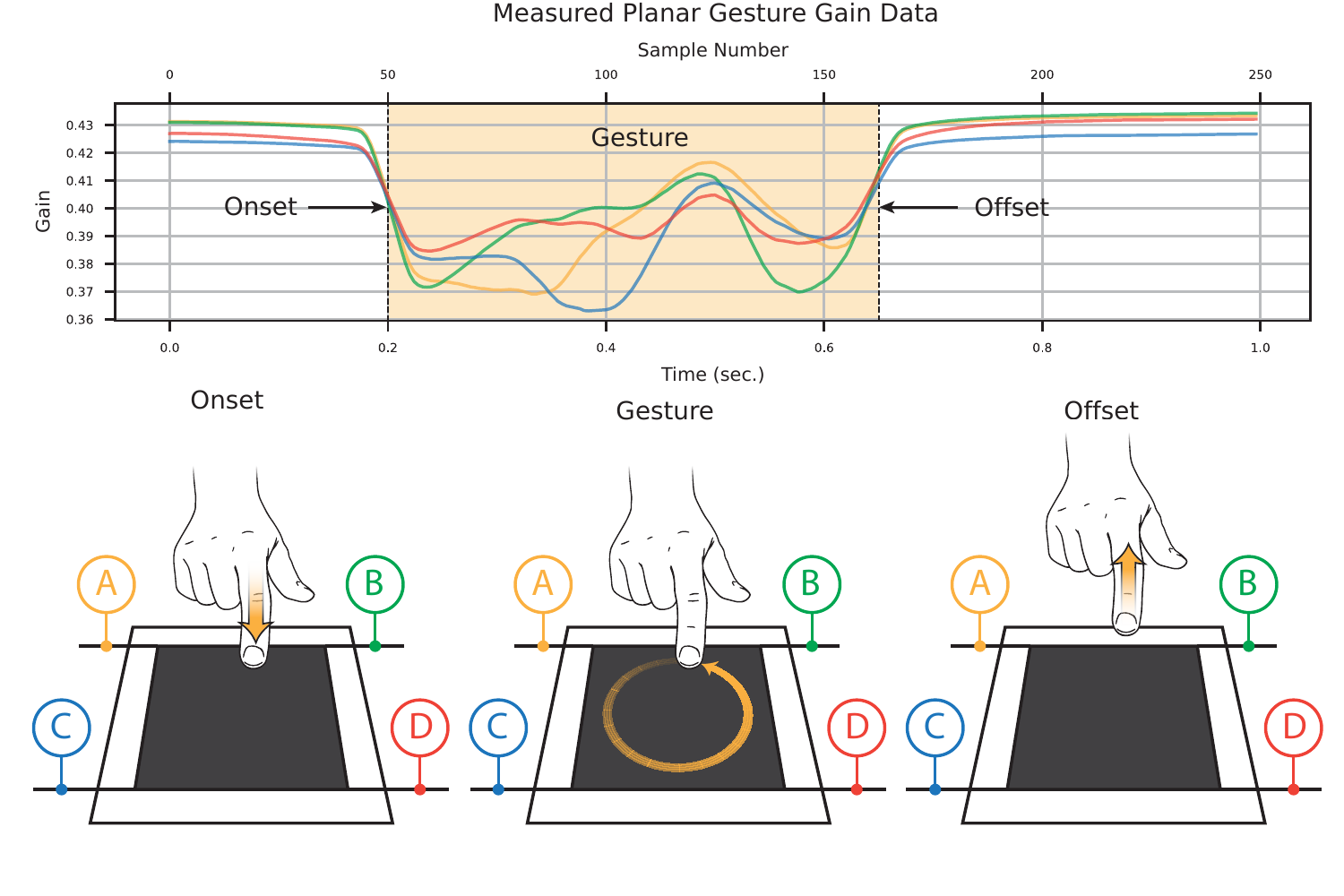}\label{fig:Planar_CTS_Gesture_Data_Example}}
    \caption{Illustrations of the measurement circuit and example recorded gesture data. \emph{(a)}: Illustration of the measurement circuit and resistor-capacitor network circuit formed during touch. \emph{(b)}: Plot of example measured data showing the changes in gain measurement during onset, gesture, and offset.}
    \Description{This figure illustrates the measurement circuit in (a), and the signal during different stages of a gesture event. In (a), an illustrated waveform generator, a digital oscilloscope, and a knitted sensor with four connection points are shown. Additionally, there is a 5-point star circuit illustrated, which includes the four sensor connections points A, B, C, D, and a user’s point of touch on the sensor. In (b), we see a graph on top with time ranging from 0 to 1 seconds on the X axis, and voltage gain values on the Y axis. There are four lines in the graph, each representing the signal measured from one of the electrodes. The plot is split into three areas in the time axis: before the gesture event, the gesture event, and after the gesture event. During the first and third phases, the behavior of the four lines is very similar, and their voltage gain values are high. However, during the gesture event, we notice a drop in the voltage gain value for all these values, and the differences among the individual lines becoming more pronounced. Below the plot, there are three illustrations of the four-connection knitted sensor with a hand touching the sensor. In the first illustration, we see the onset of touch, in the second one, the gesture being performed, and in the third, the offset.}
  \label{fig:Planar_CTS_Gesture_Data_Example}
\end{figure}

The knitted sensor design we introduce is a continuous conductive rectangular sensing area, constructed using one conductive yarn and four electrode connections points, illustrated in Figure~\ref{fig:Planar_Touchpad_Diagram}. The yarn used is outwardly conductive, since it is composed of carbon-suffused nylon, and forms a resistive mesh when knitted. The inter-row yarn loop connections form an approximate uniform resistance gradient across the planar surface. This property is beneficial to measuring touch gesture as the resistance gradient is uniform in all directions. Touch location and capacitance magnitude are inferred through voltage measurements recorded at the corner locations. Capacitive touch draws current from the current sources attached at the corners, which affect the differential charge of the circuit (Figure~\ref{fig:Planar_CTS_Touch_Sensing_Circuit}).

Patterns developed in previous work~\cite{Vallett2016a,Vallett2019a,mcdonald2020knitted}, as seen in Figure~\ref{fig:Single-wire_Touchpad_Diagram}, utilized a single linear conductive pathway that covered a planar area while following the serpentine pathway of weft knitting. In those sensors, location was inferred as a linear distance along the pathway mapped along a 2-dimensional surface. Their circuit required two connections at the endpoints of the pathway (A and B). This design strategy reduces the dimension of the measured voltage outputs to what is required to decouple linear touch location and capacitance, and improves the simplicity of connecting the textile to measurement hardware. This strategy is acceptable for inferring static touch location or tap gestures, but it is non-ideal for inferring complex gesture pathways across the surface of the sensor. The measured output response is non-uniform between geometrically adjacent touch-points which complicates precise localization. Additionally, touch-points are spaced apart into a sparse sensitive area to prevent shunting when two or more touch-points are contacted. This spacing induces loss of contact when touch is transferred to another location, thus breaking the gesture’s continuity.

\subsection{Measurement Circuit}
\label{sec:measurement_circuit}
The capacitive touch circuit formed by the planar conductive area is modeled as a mesh resistor-capacitor ladder network. Figure~\ref{fig:Planar_CTS_Touch_Sensing_Circuit} illustrates the circuit diagrams of the resistances formed between the location of touch and the sensing points ($A, B, C, D$), represented as a star graph. The values of the resistances vary depending on touch location. The value $C_{t}$ represents the magnitude of capacitance induced by touch, which is used as a pseudo-pressure. The voltage measurement locations ($A, B, C, D$) have associated parasitic capacitances ($C_{p_{A}}$, $C_{p_{B}}$, $C_{p_{C}}$, $C_{p_{D}}$) which affect the voltages at the measurement locations. The excitation waveform passes through current-limiting resistors ($R_{A}$, $R_{B}$, $R_{C}$, $R_{D}$) that reduce the current entering and exiting the fabric, which allows voltage measurements to be discerned.

An example gesture and processed measurement are shown in Figure~\ref{fig:Planar_CTS_Gesture_Data_Example}. Figure~\ref{fig:Measured_Data_Processing_Pipeline} further explains the waveform processing steps. All measured gestures consist of a distinct touch onset, gesture action, and touch offset. During onset, the measured waveform gains decrease due to the increase in touch capacitance. The gain at each measurement point attenuates as a function of touch location, where attenuation increases as touch proximity increases. This phenomenon can be seen during the gesture action, where the motion trajectory and gain attenuation pf each sensing point correlate. Once the gesture is complete, the offset action returns the gain measurements to a baseline.

Waveform generation and acquisition are performed using a Keysight 33622A arbitrary waveform generator and a Keysight MSO-X-3024T oscilloscope. The waveform generator produces a 2 MHz sine wave used as input to the circuit (output 1) and 250 Hz square wave used to trigger oscilloscope sampling (output 2). The sine wave passes through current-limiting resistors attached to the four corners of the fabric circuit. The resistance values are approximately equal to the touchpad resistance of 4 kOhm/sq. A parasitic capacitance of approximately 60 pF is observed at each measurement point. Gesture data is collected over the span of 1 second at a rate of 250 frames per second. Each frame window consists of approximately 4000 samples collected at 625 MSamples/s. The frames are processed using Bode analysis to return the magnitude ratio (gain) of the input and measured waveforms. The output data is of dimension $250\times 4$ corresponding to the 250 measurement frames and 4 measurement channels.

\subsection{Signal Filtering}
\label{sec:wavelet_filtering}

After the data is captured by the four electrodes as changes of gain values over time, we subtract a baseline event, captured while no gesture was being performed on the knitted touchpad. The purpose of baseline subtraction is to remove any elements of the representation that stay the same across different measurements, therefore highlighting the differences between gestures, which makes classification easier. 

\begin{figure}
    \centering
    \includegraphics[width=0.95\textwidth]{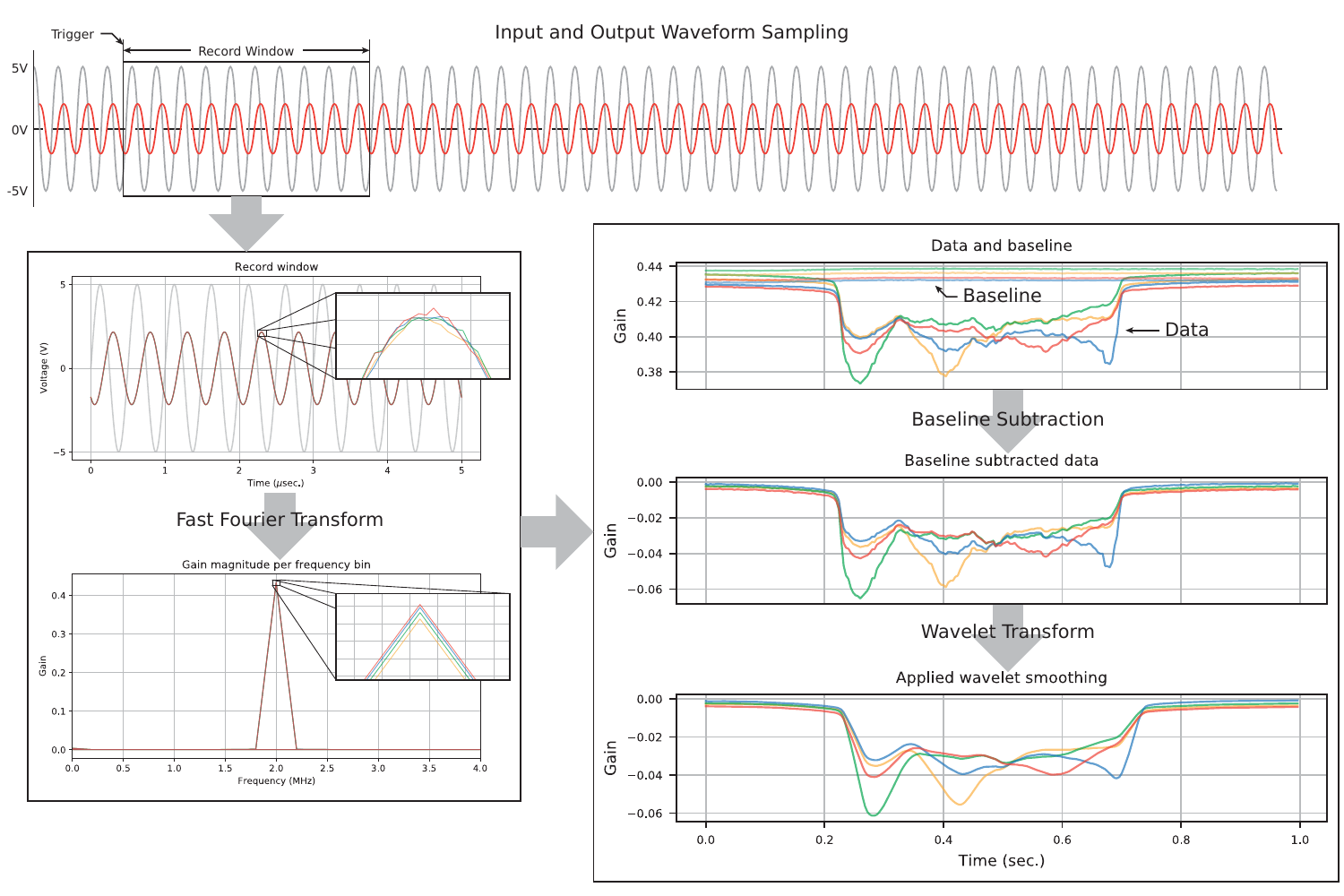}
    \caption{Diagram of the data processing pipeline: The waveform data is sampled in segmented record windows over the duration of the gesture. The FFT of the window is processed and the magnitude of the bin at the input frequency is used to determine the window gain. The measured baseline gain is subtracted from the gesture gain data and a wavelet transform is applied to smooth high-frequency changes.}
    \label{fig:Measured_Data_Processing_Pipeline}
    \Description{This figure illustrates the signal processing steps after its acquisition. There are three major steps: the first one is recording the event window, while the input and output waveforms are continuously sampled. The second is converting the recorded time-domain window into frequency domain via Fast Fourier Transform. Subsequently, the baseline value is subtracted, and then the signal is filtered through wavelet transform.}
\end{figure}

Subsequently, in order to further reduce noise from the signal, we filter it using \emph{wavelet transform}, a powerful and flexible analytic tool which allows us to obtain signal information from both its time and frequency domains. It does not only indicate which frequencies are present in the signal, like Fourier Transform, but also when those frequencies occur in time. Wavelet Transform achieves this by calculated compromises: by low frequencies having a high resolution in the frequency domain, and low resolution in the time domain; and high frequencies having a low resolution in the frequency domain, and high resolution in the time domain. Wavelet Transform analyses the signal in different scales: first, working with larger windows of the signal for elements stretched in time, like low frequencies. Then, after that information is acquired, we can progressively make the window of interest smaller to look for information in those scales. As we shrink the window, we lose the ability to capture low frequencies, however, that information should have been obtained in the previous step. Ultimately, we are able to use the collected information from different scales to reconstruct the signal. 

In order to filter the signal, we first deconstruct the signal through a \emph{Symlet} mother wavelet with 4 vanishing moments. After having obtained all the elements necessary for reconstruction through analysis at different levels, we reconstruct the signal using only a subset of them. Typically, the higher frequencies correspond to noise, so they are removed. For our analysis, we keep only the first 4 levels of components. These settings were selected such that noise is removed, but the filtered signal follows the original closely. Figure~\ref{fig:Measured_Data_Processing_Pipeline} illustrates an example of the original signal after the baseline has been subtracted, as explained above, together with its filtered version through wavelet reconstruction.  

\begin{figure}[h]
    \centering
    \includegraphics[width=\textwidth]{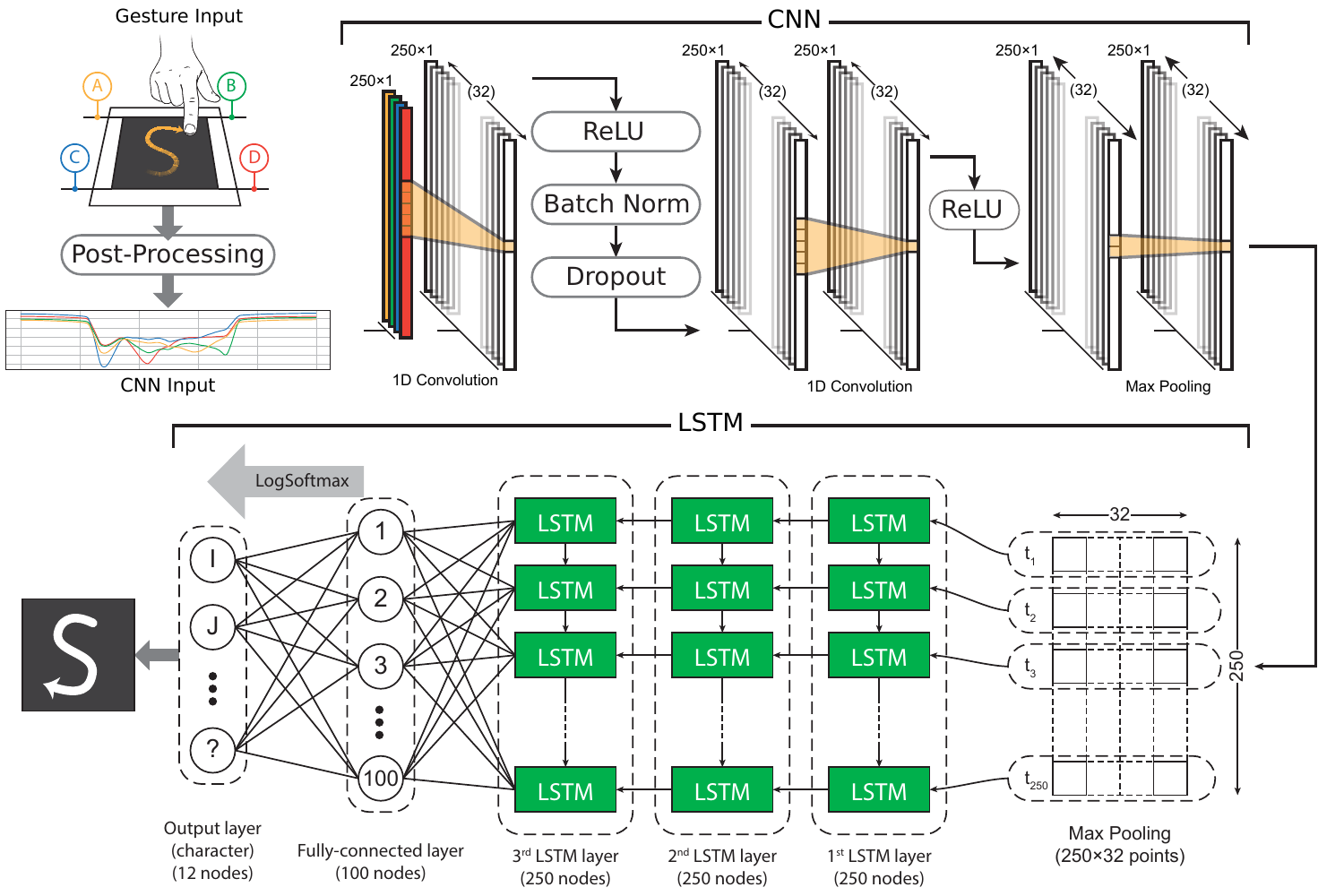}
    \caption{CNN-LSTM neural network architecture: the four-signal representation of a gesture event is used as an input after filtering, and the output is a predicted gesture category. The CNN component captures spatial relationships in the signal pattern, while the LSTM component models time-dependency.}
    \label{fig:CNN-LSTM_Network_Architecture}
    \Description{This figure shows the neural network architecture. The input to the network is the gesture performed on the sensor after it has been processed. The output is a prediction about the gesture performed, out of 12 different classes of gestures. The input first goes through the convolutional neural network, which includes functions such as convolution, max-pooling, batch normalization and dropout. Then the output of the convolutional network becomes the input to a long-short term memory network with three hidden layers of 100 nodes each. The final layer is a fully connected layer, which then maps the values to the 12-node output layer through a Log SoftMax function.}
\end{figure}

\subsection{Neural Network Architecture}
\label{sec:classification_network}
The neural network architecture used to build the recognition model relies on a combination of CNNs and LSTMs. We chose deep learning over traditional machine learning techniques based on the strengths that both CNNs and LSTMs have shown in representing time-series signal data. Additionally, these networks automatically capture the important aspects of the data without needing feature construction, typically required by most non-neural network algorithms to represent a time-series signal.

The neural network takes as an input the data after being processed as described above. The architecture, illustrated in Figure~\ref{fig:CNN-LSTM_Network_Architecture}, starts with a one-dimensional convolution layer with a rectified linear unit (ReLU) activation function, followed by a batch normalization layer and a dropout layer for regularization purposes. It continues with another one-dimensional convolution layer, again with an ReLU activation function, followed by a max-pooling layer. The function of convolution is capturing local features at different scales in the signal. Subsequently, three layers of LSTM cells follow, each with a hidden size of 100. LSTMs have been widely used for sequential data analysis, including time-series signal data, since they capture the signal's temporal dependencies. Among recurrent layer variants, they are chosen since they are capable of handling long-term dependencies in sequential data~\cite{graves2013speech,sutskever2014sequence}.

Each input to the system consists of one gesture event as captured by the four electrodes connected to the sensor, and after being transformed to filtered frequency gain values. The sequence is 250 time steps long, and there are 4 values per time step, corresponding to an input layer of size 4 to the neural network. The temporally-related frequency gain values are transformed to one of 12 possible gesture classes through a 12-output linear layer at the end of the network. In order for the network to predict a correct gesture, it needs to be trained with user data, which we discuss in Sections~\ref{sec:experiments} and~\ref{sec:results}, together with the training hyper-parameter choices.

\subsection{Model Deployment}
\label{sec:model_deployment}
For an interactive system, after the above architecture is used for training, the resulting model needs to be deployed to lightweight hardware. We use a NVIDIA® Jetson Xavier™ NX Developer Kit  (Figure~\ref{fig:nvidia_picture}), running Ubuntu 18.04 with 8GB of RAM, a 6-core NVIDIA Carmel ARM CPU, and a NVIDIA Volta GPU for gesture classification (Table~\ref{tab:nvidia_specs}). The NVIDIA board's ability to do fast parallel math is important, as it allows running pre-trained models. In addition, its form factor is relatively compact, an important feature for many of the potential applications of the sensor.

Currently, the data used as the model input is a \emph{.csv} file containing one user-captured touchpad gesture. First, we subtract the baseline readings and perform waveform filtering to prepare the data for evaluation. Subsequently, the resulting waveform is input to the model for classification. The model then returns its prediction as an output, which is one of the 12 gesture classes. 

In the future, we plan to adapt the model deployment to handle continuous or real-time signal classification, as opposed to only pre-recorded discrete signals. A barrier to real-time signal classification is the practical lack of hardware that can perform the signal generation, acquisition, and filtering at the speeds necessary to continuously supply the model with new inputs. Custom hardware, which we aim to design, would be able to stream the acquired signal to the NVIDIA Jetson board for classification. Once the Jetson board would acquire the signal, it would need to parse it to isolate individual gesture windows, since the model would have been trained on inputs containing only a single gesture. An additional model may be necessary to accurately distinguish gestures. The main gesture classification model could then take gestures as inputs as they are returned from the parser.

\vspace{4mm}
\noindent
\begin{minipage}{\textwidth}
  \begin{minipage}[b]{0.49\textwidth}
      \centering
      \includegraphics[width=0.7\linewidth]{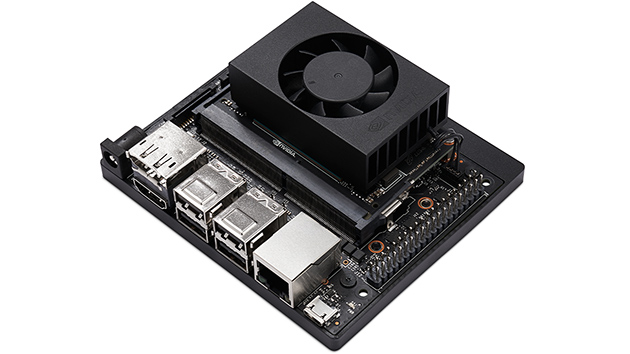}
      \captionof{figure}{NVIDIA Jetson Xavier NX Developer Kit~\cite{jetsonxaviernx}}
      \label{fig:nvidia_picture}
  \end{minipage}
  \hfill
  \begin{minipage}[b]{0.49\textwidth}
    \centering
    \captionof{table}{Hardware Specifications}
    \label{tab:nvidia_specs}
    \begin{tabular}[b]{c|p{0.7\textwidth}}
        \small{GPU} & NVIDIA Volta Architecture with 384 CUDA cores and 48 Tensor cores \\
        \hline
        \small{CPU}  &  6-core NVIDIA Carmel ARM Processor \\
        \hline
        \small{RAM} & 8GB LPDDR4x \\
        \hline
        \small{Size} & 70 mm x 45 mm\\
    \end{tabular}
    
  \end{minipage}
\end{minipage}
\vspace{4mm}

\section{Experiments}
\label{sec:experiments}
\begin{figure*}[h]
    \centering
    \subfigure[]{\raisebox{3mm}
    {\includegraphics[width=0.22\textwidth]{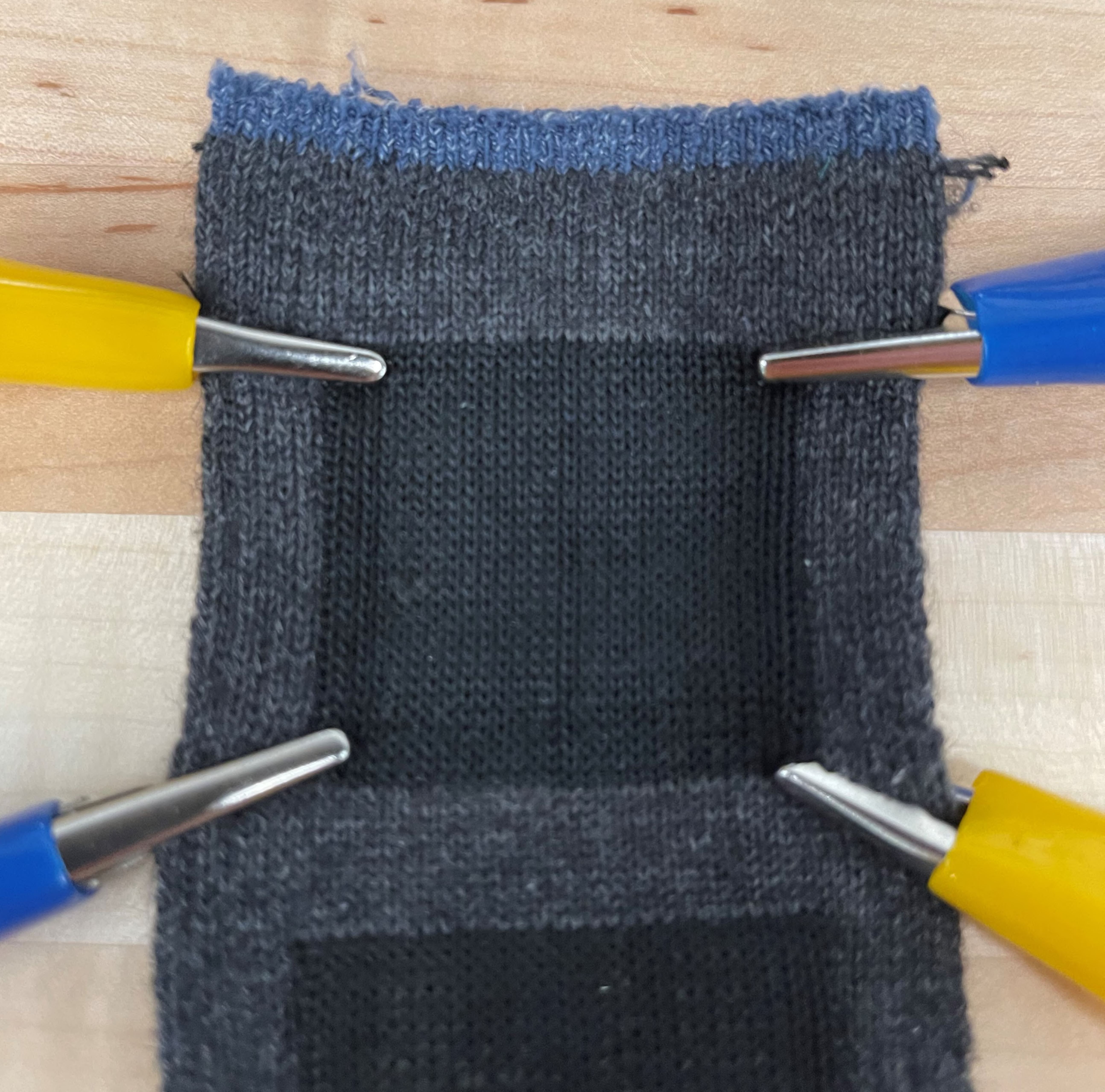} \label{fig:capturing_gesture}}}
    \hfill
    \subfigure[]
    {\includegraphics[width=0.7\textwidth]{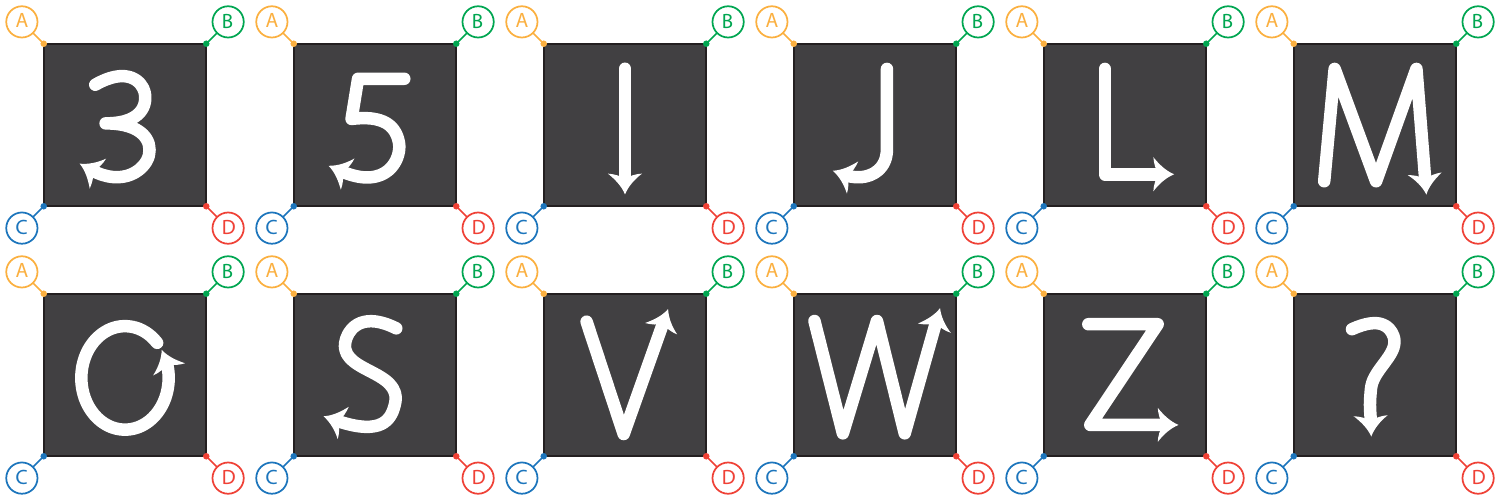} \label{fig:gesture_vis}}
    \caption[Gesture pathways]{Pathways of the collected gestures. \emph{(a):} The knitted component of the sensor on which the gestures were performed; \emph{(b):} All the gesture pathways that were collected and analysed.}
    \label{fig:Gesture_Pathways}
    \Description{This figure provides information regarding the data collection process. In (a), there is a picture of the knitted sensor with four connected electrodes. Figure (b) shows the pathways for all the 12 types of gestures performed. They are the English language characters: ‘3’, ‘5’, ‘I’, ‘J’, ‘L’, ‘M’, ‘O’, ‘S’, ‘V’, ‘W’, ‘Z’, ‘?’. Each sketch shows the shape of the gesture with an arrow pointing outward at the end of the gesture to clarify the direction of the performance.}
\end{figure*}

In order to train the neural network models, we rely on user data, collected using the circuit described in Section~\ref{sec:measurement_circuit}. We used a touchpad sensor like the one shown in Figure~\ref{fig:capturing_gesture}, connected to a Keysight 66322A Series Waveform Generator and a Keysight InfiniiVision MSO-X 3024T mixed signal Oscilloscope, for signal generation and acquisition respectively. A group of 8 total college-age subjects participated in data collection. All of the participants were in good health. For a single trial, participants drew one of the 12 gestures included in the study onto the touchpad sensor, as the oscilloscope was capturing the resulting waveform. The gesture pathways captured are illustrated in Figure~\ref{fig:gesture_vis}. For each gesture, a baseline reading was also captured without any input to the touchpad, to be further used for data processing as described in Section~\ref{sec:wavelet_filtering}. A few of the subjects completed trials over different sittings. In those cases, a baseline measurement was taken for each sitting to account for outside conditions. 

Training was accomplished using leave-one-out cross validation for each of 5 subjects. Additionally, we evaluated the performance of the trained models, by testing their accuracies against data from 3 new subjects, to further investigate the robustness and reliability of the results. Each participant performed at least 20 trials for each of the 12 gestures included in the experiment, with an average of $45$ samples. A total of $2700$ samples were collected for cross-validation, with an average of $225$ samples per class. The evaluation set was composed of a total of $720$ samples, or $60$ per class, with each subject performing $20$ trials per gesture type. We investigate whether our gesture recognition model is capable of accurately distinguishing among 12 different language character gestures. The collected and processed data is used as an input into the CNN-LSTM gesture recognition network, which outputs one of the 12 possible gestures. Accuracy, precision, recall, and F1-score are used to determine its performance. More details about the process of training and evaluation are provided in the section below.

\newsavebox{\gesturebox}

\section{Methods and Results} 
\label{sec:results}

Data from 5 subjects was used for training and cross-validation, while data from 3 other subjects was used to evaluate the trained models. During cross-validation, at any time, data from 4 subjects was used for training, and data from the one other subject was used for validation. Therefore, 5 total trained models were produced, each of which was evaluated using evaluation data. Any one of those models could be deployed for use in a real-time interactive system.

The subject data was acquired as described in Section~\ref{sec:experiments}, and processed as detailed in Section~\ref{sec:wavelet_filtering}. Before training, we balanced the data by oversampling within each subject, such that each class had the same number of samples, helping to achieve model stability. The balanced training dataset resulted in an average of $277$ samples per class, while the evaluation dataset was already balanced at $60$ samples per class. Our model, whose architecture was defined in Section~\ref{sec:classification_network}, was trained using PyTorch 1.4 \cite{paszke2017automatic} with CUDA 10.0, using a learning rate $\alpha = 0.001$, a dropout rate = $0.6$ and a batch size $= 128$, over $2000$ epochs. We used \emph{Adam} as an optimizer, and a negative log likelihood loss function. The network weights were initialized using Xavier initialization~\cite{glorot2010understanding}.

These hyper-parameters were the same as the ones in~\cite{mcdonald2020knitted}, where an LSTM-based neural network model was used to identify location of touch on a 36-button knitted sensor. In addition to using the subject data to train our CNN-LSTM gesture recognition model, we also use it to train an LSTM-based model as a baseline comparison. The only difference between the two methods is the neural network architecture. The technique used in in~\cite{mcdonald2020knitted} also involves a statistical feature construction step over the 92 analyzed frequency gains, which was not possible to apply to our current data, since, due to hardware constraints, only relies on a single frequency. The results of using the baseline architecture are included in Table~\ref{tab:results}, together with those of our neural network model.   

\subsection{Cross-Validation and Evaluation Results}

Our model's cross-validation \emph{accuracy}, reported in Table~\ref{tab:results}, is 78.6\%. This accuracy was calculated as the average accuracy of the 5 cross-validation folds. This means that the data from one subject at a time was tested against the model trained on the data from the other 4 subjects, and subsequently, the average of the results is reported. The accuracy for each fold was calculated as the average of the last 50 epochs. The test dataset performance, which aims to capture how generalizable the model is, shows an even greater accuracy of 88.5\%. This accuracy was calculated as the average of testing the 3 test subject data against the 5 trained models produced during cross-validation. 

\begin{table}[h]
  \centering
  \caption{Classification results for model cross-validation and evaluation.}~\label{tab:results}
  \vspace{0.5cm}
  \begin{tabular}{l|cccc|cccc}
  \vspace{3mm}
    & \multicolumn{4}{c}{\small \textit{Cross-Validation}} & \multicolumn{4}{c}{\small \textit{Evaluation}} \\
    
    {\small\textit{Method}} & {\small \textit{Accuracy}} & {\small \textit{F1-Score}} & {\small \textit{Precision}} & {\small \textit{Recall}} & {\small \textit{Accuracy}} & {\small \textit{F1-Score}} & {\small \textit{Precision}} & {\small \textit{Recall}}\\
    
    \midrule
    \small{LSTM}         & 41.7\% & 22.4\% & 25.4\% & 22.0\% & 52.5\% & 53.9\% & 61.6\% & 52.5\% \\
    \small{CNN + LSTM}   & 78.6\% & 73.5\% & 73.9\% & 73.7\% & 89.8\% & 89.7\% & 90.3\% & 89.8\% \\
  \end{tabular}
  \Description{This table shows the cross-validation and evaluation performance measures for the baseline method, which is a long-short term memory network (LSTM), and for the method proposed in this work, a combination of convolutional and long-short term memory network (CNN-LSTM). The performance measures are accuracy, precision, recall, and F1-score. The CNN-LSTM network significantly outperforms the LSTM one in all measure, resulting in an accuracy of 78.6\% for cross-validation and 89.8\% for evaluation.}
\end{table}

\begin{figure*}[ht]
    \centering
    \subfigure[]
    {\includegraphics[width=0.495\textwidth]{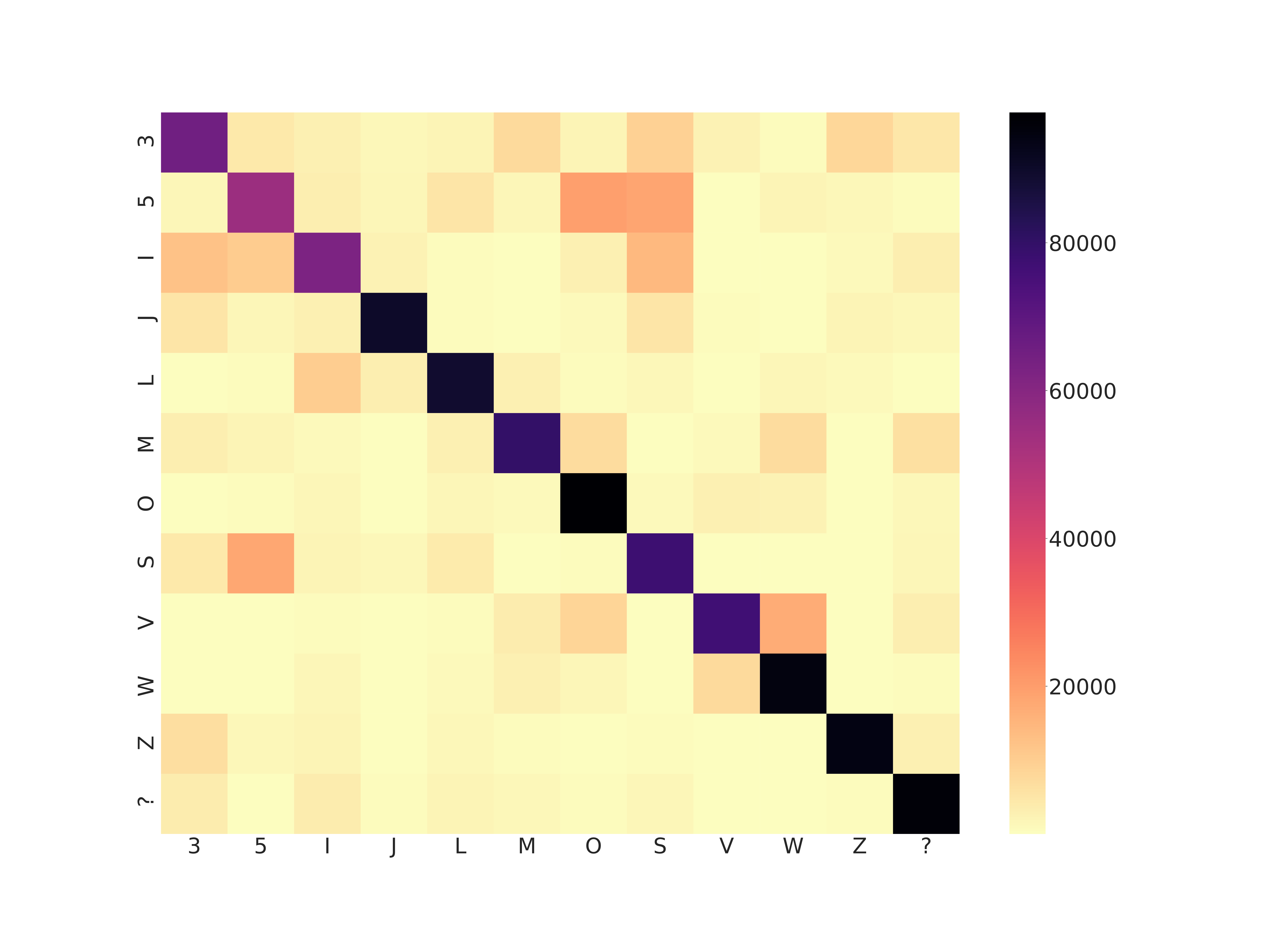}\label{fig:cv_confusion}}
    \hfill
    \subfigure[]
    {\includegraphics[width=0.495\textwidth]{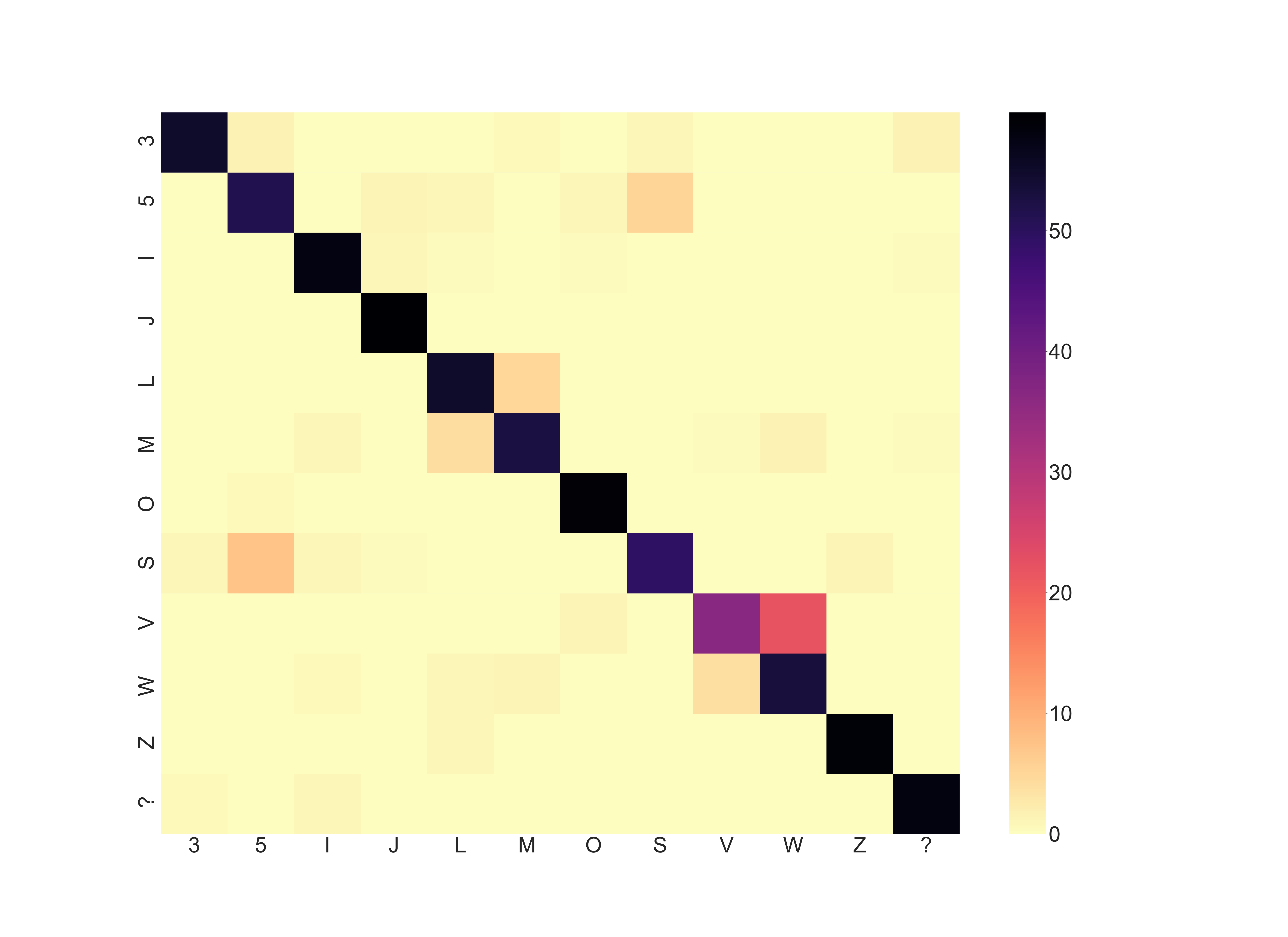}\label{fig:test_confusion}}
    \hfill
    \caption{Classification matrices produced with results from cross-validation \emph{(a)}, and evaluation \emph{(b)}. The matrix rows denote the true row categories of the gestures, while the columns show the ones predicted during evaluation. Each category is denoted by the respective gesture pathway performed on the knitted touchpad. A higher-value diagonal, with the rest of the matrix having lower values, would be a desirable combination indicating a high accuracy. This figure illustrates how most gestures are correctly classified - if gestures are similar to each-other however, it is easy to incorrectly classify them as each-other.}
    \label{fig:Confusion_Matrices}
    \Description{This figure shows two classification matrices, one for the results of cross-validation in (a), and the other for the results of evaluation in (b). Both the rows and the columns of each matrix are labeled by the gesture types; therefore, the size of each matrix is 12x12. In both cases the diagonal along each is a much darker color (dark purple to black) compared to the surrounding values (yellow to orange).}
\end{figure*}

Other measures of performance we calculate are \emph{macro-precision, macro-recall}, and \emph{macro-F1-score}. Precision refers to the ratio of true positives to the combined number of true and false positives. Recall is the ratio of true positives to the combined true positives and false negatives. The F1-score is the harmonic mean of precision and recall: $F1 = 2 * (precision * recall) / (precision + recall)$. Macro-precision, macro-recall, and macro-F1 refer to the balanced respective scores per class, and those values are obtained by averaging all respective class scores. In Table~\ref{tab:results} these values are simply referred to as: \emph{precision}, \emph{recall}, and \emph{F1-score}. These scores give greater insight into how specific accuracies were achieved. Additionally, we provide the classification matrices of all gestures in Figure~\ref{fig:Confusion_Matrices}, for both cross-validation and evaluation of our CNN-LSTM model.

As seen from Figure~\ref{fig:Confusion_Matrices}, there are gestures that incorrectly get classified as others, with pairs such as \emph{'S'} and \emph{'5'}, or \emph{'V'} and \emph{'W'} being noticeable. This can be expected as the trajectories between the two gestures are very similar. It should be noted that these results are not rotation-invariant, due to the nature of the gestures selected. Otherwise, the users' intention regarding entering a \emph{M} versus a \emph{W} would be obscured. Overall however, the model accurately identifies gestures performed on it, and generalizes very well, most probably in part, because the architecture includes batch normalization and a relatively high dropout rate of 0.6. If we compare the performance of this model to the baseline, we notice that the CNN-LSTM model significantly out-performs the baseline in all measures. 

The accuracies reported here are subject-independent, even though subject-specific calibration is expected to increase overall accuracy. If a knitted pattern is designed to be used by a single user, calibration is likely to increase the gesture detection accuracy, since it removes variations that are caused by different users' physiological states, or specific ways of performing gestures. However, designing a system that does not rely on calibration makes the experience of multiple users interacting with the same sensor much more intuitive and non-intrusive. Additionally, we expect that provided more data is available, the model's accuracy would increase, since neural network models typically require a large amount of data to produce accurate results. Moreover, training under a variety of conditions, such as different environmental condition and other possible distortions, would mean that the model would incorporate the signal variations induced, and be more robust when exposed to them.   

\subsection{Resources and Time Performance}
The machine on which our models were trained through cross-validation, and later evaluated was running Ubuntu 18.04, with an Intel\textsuperscript{\textregistered}~Core\textsuperscript{\texttrademark} i9-9960X CPU @ 3.10GHz, 128 GB RAM, and 2x RTX 2080 Ti Blowers with NVLink cards. The trained models were deployed and evaluated on a NVIDIA Jetson Xavier NX Development Kit, described in Section~\ref{sec:model_deployment}, which was also running Ubuntu 18.04. We measured the total runtime of 3 different stages of the evaluation process: the time to convert the input from a numpy array to a CUDA tensor, the time taken for the loaded model to evaluate the input tensor, and the total time taken for one trial including the first two measurements. The results of these measurements are included in Table~\ref{tab:timing}.

\begin{table}[ht]
    \centering
    \caption{Timing measurements of model execution}~\label{tab:timing}
    \begin{tabular}{l|ccc}
         &  \small \textit{Conversion to CUDA} & \small \textit{Sample Evaluation} & \small \textit{Total Trial}\\
         \midrule
        \small \textit{Bootstrap and First Trial Time} & 3.7s & 3.6s & 7.3s\\
        \small \textit{Avg. Trial Time after first} & $1.9\times10^{-4}$s & $1.7\times10^{-2}$s & $3.0\times10^{-2}$s\\
        \Description{This table shows the time response for processing the first sample, as well as the average of several samples after the first, measured while running on the NVIDIA® Jetson Xavier NX™ Developer Kit. There are three columns to it: ‘conversion to CUDA’, ‘sample evaluation’, and ‘total trial’. The total trial time is 30ms for an average trial.}
    \end{tabular}
\end{table}

The purpose of this evaluation was to determine the feasibility of building a responsive real-time interactive system given a trained gesture recognition model. From the results in Table~\ref{tab:timing}, we can see that the average response time of a typical evaluation is approximately 30 ms, a very reasonable response time for real-time systems. The first trial's response time is much greater since it involves bootstrap and the NVIDIA CUDA® Deep Neural Network library (cuDNN) performing cache allocations. Typically, that first trial in real-world applications can be substituted with a sample that is not relevant to the gesture signal evaluation. For a fully-interactive system, the response time of the signal acquisition and processing controller would need to be added to that response time. Even though that component is not implemented in this work, prior work~\cite{mcdonald2020knitted} has shown that it is possible to have a responsive micro-controller performing those functionalities. Such systems could enable many applications, some of which are discussed in the following section, together with other considerations for real-world interactivity.

\section{Real-World Applicability}
The experimental results in Section~\ref{sec:results} are encouraging for the feasibility of our proposed technology, however they were performed in a controlled laboratory setting. Specifically, for all those experiments the sensor pinned in a static position on the table during data acquisition. In order to expand the experimental set up of CTS gesture pads and evaluate their functionality and accurately in a setup resembling the  real world, there are many many other aspects to be considered. In this section, we begin discussing and investigating some of these possibilities, even though further explorations are necessary. First, we conduct a new study, with subjects wearing the sensor while inputting gesture data. Next, we evaluate the effect that washing and drying have on the resistance of the sensor. Finally, we discuss the application potential and integration of this system.  

\subsection{Model Performance While the Sensor is Being Worn}
\label{sec:wearing_sensor}
One of the main ways the proposed sensor is expected to be used in the real world is via clothing integration, which is facilitated by the compact sensing area. The sensor in proximity to the human body induces capacitance, which is the principle upon which its circuit design relies. When worn near the skin, there is an additional capacitance induced, greater than the baseline parasitic capacitance. Ideally, when worn, the skin should not contact the sensor. Shielding the sensor from underside contact is currently achieved by knitting a back layer on the sensor and routing conductive yarn only on the top layer. 

\begin{figure*}[ht]
    \centering
    \subfigure[]{\raisebox{5.5mm}
    {\includegraphics[height=0.285\textwidth]{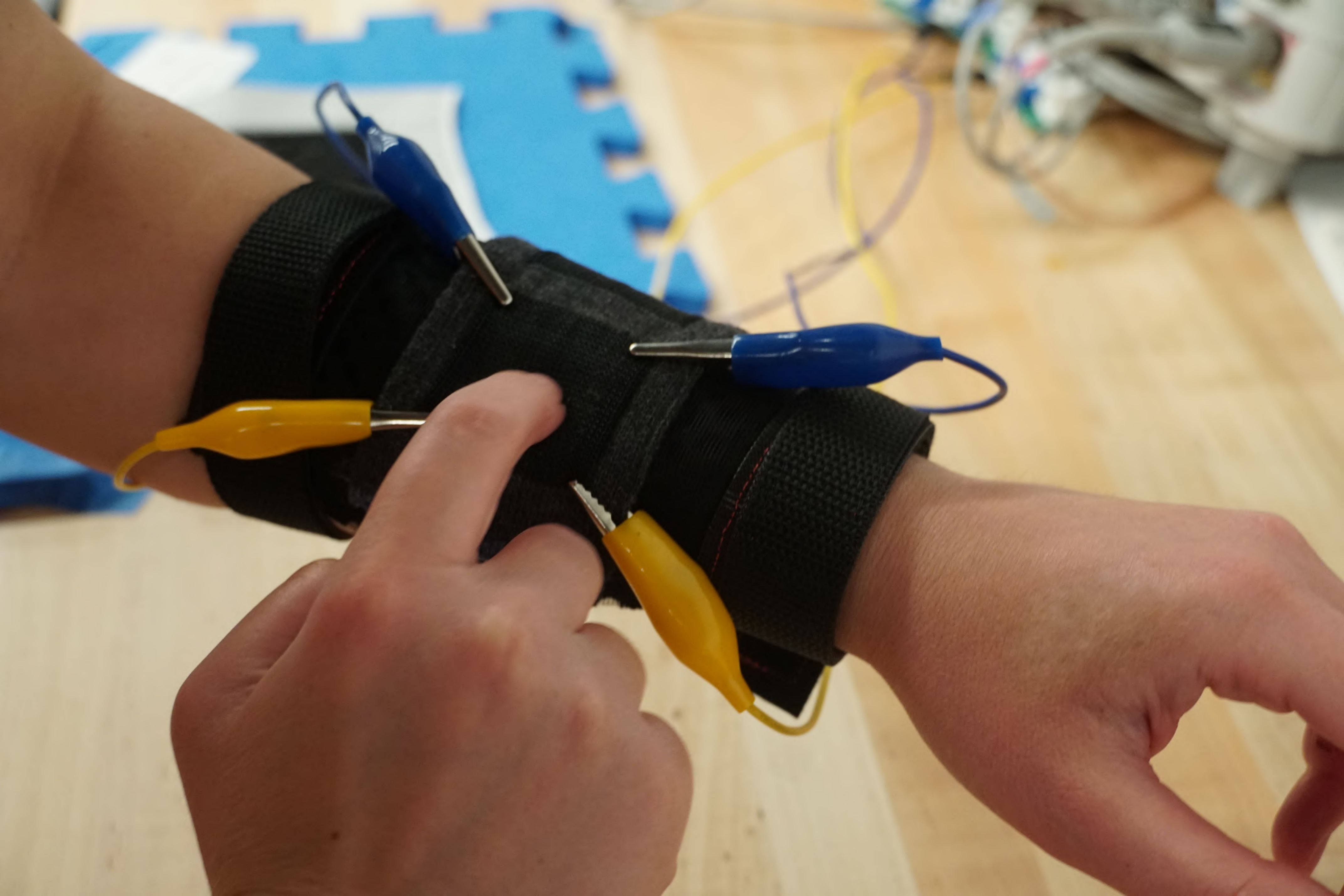}\label{fig:sensor_wearing_forearm}}}
    \hfill
    \subfigure[]
    {\includegraphics[width=0.495\textwidth]{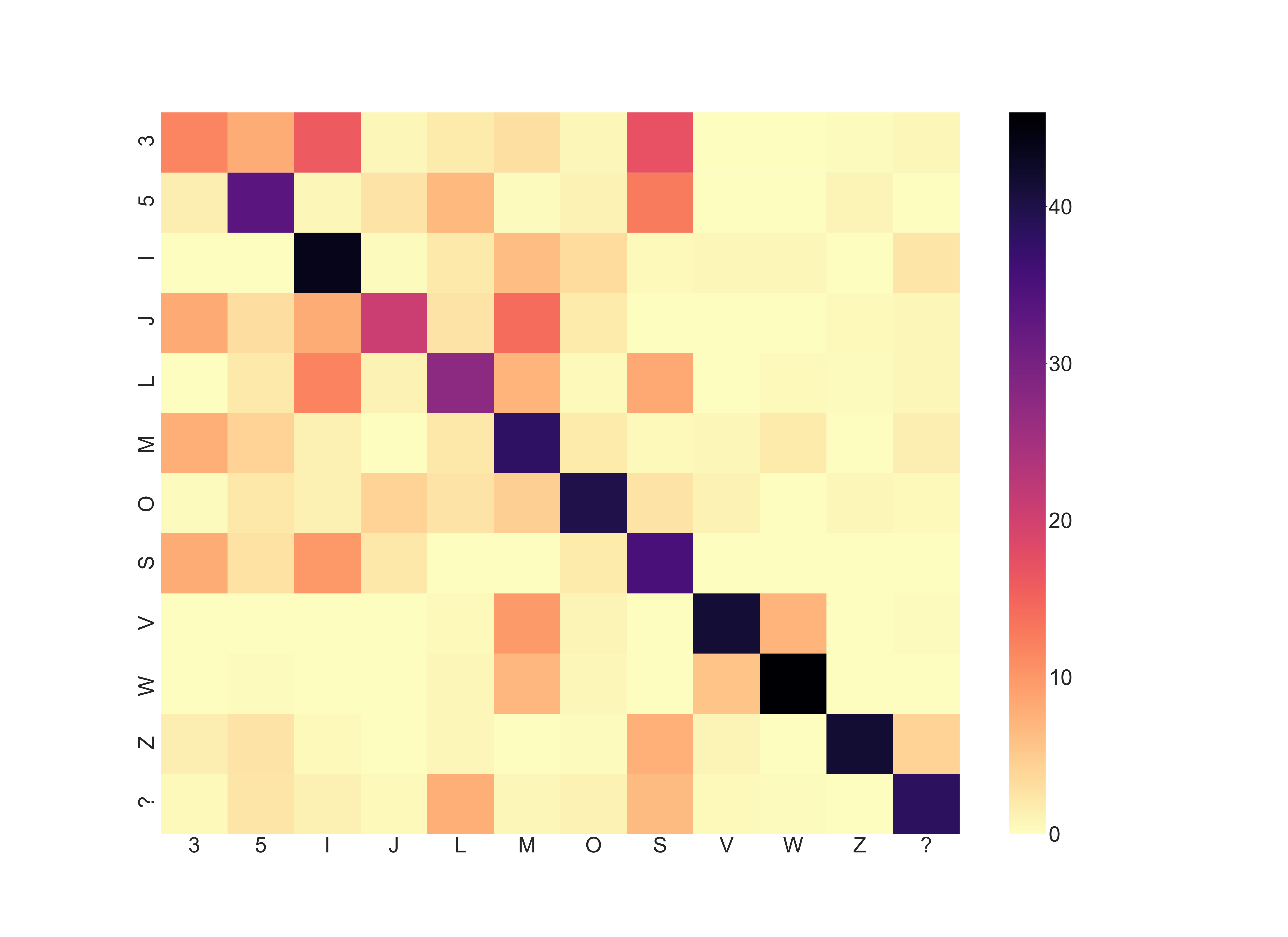}\label{fig:wearing_confusion}}
    \hfill
    \caption{User study to test model performance while knitted sensor is being worn. The setup in \emph{(a)} shows the knitted sensor fixed on a removable Velcro-strapped pad to be worn on the forearm. Four electrodes are connected to the corners of the sensing area, similarly to Figure~\ref{fig:capturing_gesture} for data collection. The heatmap generated in \emph{(b)} shows the classification results for each gesture pathway. The matrix rows denote the true row categories of the gestures, while the columns show the ones predicted during evaluation. There is a clear difference between this figure and the heatmaps in Figure~\ref{fig:Confusion_Matrices}. However the diagonal in the middle is still distinguishable from the rest of the values.}
    \label{fig:wearing_sensor}
    \Description{This figure shows a user touching on a knitted sensor while wearing it on their forearm in (a). Figure (b) shows the gesture classification matrix, with the diagonal values being higher than the rest of the values, but less pronounced than the diagonals of Figure 10.}
\end{figure*}

\subsubsection{Methods and Results}
In order to further examine the robustness of the gesture recognizing model introduced above and its usefulness in the real world, we conducted a new user study with 3 subjects. The study is similar in design and analysis to the cross-validation and evaluation studies described in Sections~\ref{sec:experiments} and~\ref{sec:results}, but in this case, the subjects were wearing the sensor on their forearm while performing the same 12 gestures as in the studies above. The sensor was pinned on another Velcro-strapped pad as shown in Figure~\ref{fig:sensor_wearing_forearm}. Subjects were not moving while collecting the data, but some motion of their arms would be expected. Each subject performed each of the 12 gestures 20 times, with a total of 720 samples, or 60 per class collected. This dataset was tested against the already trained models from the cross-validation study. No additional training was performed to account for this new condition.

\begin{table}[h]
  \centering
  \caption{Classification results for gestures performed while the sensor was being worn.}~\label{tab:wearing_results}
  \vspace{0.5cm}
  \begin{tabular}{l|cccc}
    & {\small \textit{Accuracy}} & {\small \textit{F1-Score}} & {\small \textit{Precision}} & {\small \textit{Recall}}\\
    \midrule
    \small{CNN + LSTM} & 58.0\% & 58.2\% & 61.2\% & 58.0\% \\
    \Description{This table shows the evaluation performance measures of the sensor while it is being worn using trained CNN-LSTM model. The performance measures are accuracy, precision, recall, and F1-score. The evaluation accuracy is 58.0\%.}
  \end{tabular}
\end{table}

Table~\ref{tab:wearing_results} demonstrates the results computed for the same performance measures as the studies above, including: \emph{accuracy}, \emph{precision}, \emph{recall}, and \emph{F1-score}. Figure~\ref{fig:wearing_confusion} illustrates the classification matrix of the real and predicted gestures. The computed accuracy is $58.0\%$, which is lower than both the cross-validation and testing accuracy, however well above chance accuracy (8.3\%). We would expect increased accuracy with training that includes samples collected while the sensor is being worn. For a complete system implementation, training should be performed under a large variety of conditions to ensure robustness to different real-world circumstances.

\subsection{Effect of Washing and Drying on Sensor Resistance}
\label{sec:washing_drying}
\begin{wrapfigure}{l}{0.35\linewidth}
  \centering
    \includegraphics[width=0.8\linewidth]{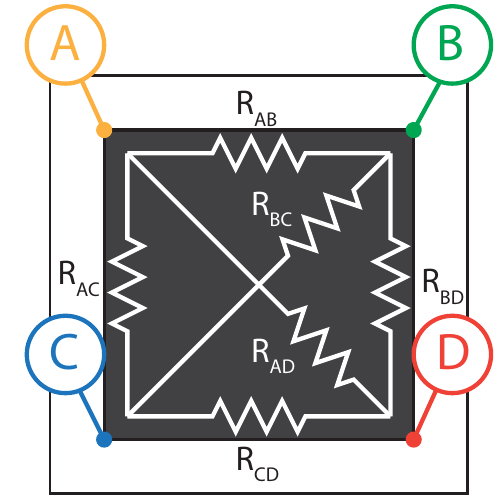}
    \captionof{figure}{Annotated sketch of the knitted sensor, showing points along which resistance was measured.}~\label{fig:sensor_sketch}
    \Description{This figure shows the sketch of the knitted sensor, illustrating four electrode connection points, A, B, C, D, each in one corner of the rectangular area. The edges and diagonals of the figure are annotated to show a resistance between each each pair of points.}
\end{wrapfigure}

Another aspect in evaluating the robustness for use of a knitted sensor is the effect of washing and drying on its conductivity. 
The resistance of the sensor is an important property in representing the resulting signal, and subsequently building a gesture-recognizing learning model~\cite{Vallett2016a,Vallett2019a,mcdonald2020knitted}. Minor changes in resistance from activities like folding and stretching are anticipated and can be accounted for within the sensing and signal processing pipeline. Furthermore, laundering is an essential post-processing step in the manufacturing process that permanently sets physical yarn properties, such as expanding heat-bulking Nylon fibers, which in turn alter the baseline electrical conductivity. The sample tested has been washed and dried before these experiments, in addition to being steamed, as part of its manufacturing process. In this experiment, we first measure the baseline resistance across every pair of connection points illustrated in Figure~\ref{fig:sensor_sketch}. Then, we wash and dry the sensor for five cycles, measuring the resistance across the same pairs of points after each cycle.

Planar conductivity involving multiple connection points is described using the symmetric matrix shown in (\ref{eqn:Conductivity_Matrix}), which is an extrapolation of \emph{Kirchhoff's Current Law} stating that the sum of the currents entering a node is equivalent to the sum of the currents exiting it. In this application, conductivity, $G$, is directly proportional to current, $I$, and inversely proportional to resistance, $R$, such that $I \simeq G=R^{-1}$. The inverse of the resistance values indicated in Figure~\ref{fig:sensor_sketch} comprise the non-diagonal elements of the conductivity matrix. The change in conductivity is assumed to be scalar in that the values of the conductivity matrix will change proportionally. In practice, these values vary due to local changes in conductivity. Therefore, the average change in values is used to describe the cumulative change in conductivity.

\small
\begin{equation}
    G_{ABCD} = 
    \begin{bmatrix}
        -\left(G_{AB} + G_{AC} + G_{AD}\right) && G_{AB} && G_{AC} && G_{AD} \\
        G_{AB} && -\left(G_{AB} + G_{BC} + G_{BD}\right) && G_{BC} && G_{BD} \\
        G_{AC} && G_{BC} && -\left(G_{BC} + G_{AC} + G_{CD}\right) && G_{CD} \\
         G_{AD} && G_{BD} && G_{CD} && -\left(G_{AD} + G_{BD} + G_{CD}\right)
    \end{bmatrix}
    \label{eqn:Conductivity_Matrix}
\end{equation}

\normalsize

\subsubsection{Procedure and Results}
The sensor was washed according to the American Association of Textile Chemists and Colorists (AATCC) Laboratory Procedure 1-2018 Home Laundering: Machine Washing protocol~\cite{AATCC}. This protocol specifies a 35 minute wash duration with $1.8$ kg of laundry and $66 \pm 1$ g of detergent, and a standard tumble drying protocol with a temperature of $68 \pm 6^\circ C$. This protocol was chosen as appropriate for everyday laundering of clothing.

For the proposed sensor, we measured 6 resistance values for each of the two conditions: \emph{baseline ($b$)}, and \emph{washing and drying ($d_n$)}, where $n =$ 1 to 5 indicates the cycle number. The values were measured between every pair of corners in the sensor, annotated as $A, B, C$ and $D$ in Figure~\ref{fig:sensor_sketch}, which represent the connection points to the measurement hardware. To fully characterize the resistance across the conductive area of the sensor the resistance between each pair of connection points is necessary. For each resistance measurement, 100 samples were captured using a Keysight 34465A digital multi-meter, which were then averaged to represent the value of that measurement. The results are included in Table~\ref{tab:wash_dry_resistance}. For each test, the percent change in resistance between the baseline measurements and measurements after each washing and drying cycle was calculated as $\%\Delta R_{(b,d_n)} = (R_{d_n} - R_b) / R_b$. In this case, $R_{d_n}$ stands for the resistance value between the two points after the $n^{-th}$ washing and drying cycle $(d_n)$, and $R_b$ for the baseline resistance value between those same points, before any of the washing and drying cycles recorded. 
The cumulative change in resistance is calculated from the cumulative average of the element-wise matrix division of the baseline and drying cycle conductivity matrices formed using the relation in (\ref{eqn:Conductivity_Matrix}), where $\%\Delta R_{\left(b,d_{n}\right)} = avg\left(G_{b} / \left(G_{d_{n}} - G_{b}\right)\right)$.

\vspace{0.2cm}
\begin{table*}[ht]
  \caption{The percent change in resistance ($\%\Delta R$) between the \emph{baseline ($b$)} and each of the experiments after \emph{washing and drying ($d$)} is reported. The resistance values are measured between all pairs of the sensor's connection points. The cumulative change in resistance is also reported.}~\label{tab:wash_dry_resistance}
  \begin{tabular}{c|cccccc|c}
    & {\small $[A, B]$} & {\small $[A, C]$} & {\small $[A, D]$} & {\small $[B, C]$} & {\small $[B, D]$} & {\small $[C, D]$} & {\small $Cumulative$}\\
    
    \midrule
    
    \small \textit{$\%\Delta R_{(b, d1)}$} & \DeltaRBDryOneAB & \DeltaRBDryOneAC & \DeltaRBDryOneAD & \DeltaRBDryOneBC & \DeltaRBDryOneBD & \DeltaRBDryOneCD & $8.10\%$ \\
    \small \textit{$\%\Delta R_{(b, d2)}$} & \DeltaRBDryTwoAB & \DeltaRBDryTwoAC & \DeltaRBDryTwoAD & \DeltaRBDryTwoBC & \DeltaRBDryTwoBD & \DeltaRBDryTwoCD & $22.16\%$ \\
    \small \textit{$\%\Delta R_{(b, d3)}$} & \DeltaRBDryThreeAB & \DeltaRBDryThreeAC & \DeltaRBDryThreeAD & \DeltaRBDryThreeBC & \DeltaRBDryThreeBD & \DeltaRBDryThreeCD & $9.69\%$ \\
    \small \textit{$\%\Delta R_{(b, d4)}$} & \DeltaRBDryFourAB & \DeltaRBDryFourAC & \DeltaRBDryFourAD & \DeltaRBDryFourBC & \DeltaRBDryFourBD & \DeltaRBDryFourCD & $11.23\%$ \\
    \small \textit{$\%\Delta R_{(b, d5)}$} & \DeltaRBDryFiveAB & \DeltaRBDryFiveAC & \DeltaRBDryFiveAD & \DeltaRBDryFiveBC & \DeltaRBDryFiveBD & \DeltaRBDryFiveCD & $3.19\%$ \\
    \Description{This table shows the effect that washing and drying have on the resistance. The rows show the change in resistance between the baseline, and values measured after each washing and drying cycle 1-5. The columns show the point pairs along which the resistance is measured: [A,B], [A,C], [A,D], [B,C], [B,D], [C,D], and additionally the cumulative resistance change value. Most of the cumulative results values are between 3\% and 11\%, while in the rest of the table, values range from 1\% to 31\%.}
  \end{tabular}
\end{table*}

The results in Table~\ref{tab:wash_dry_resistance} show that there is stability in the resistance measurements of the sensor after washing and drying it. However, there is a change in the cumulative resistance across washing and drying trials. The fact that the change in overall resistance does not substantially increase from trial to trial is promising. However, the change in resistance varies from approximately from 1\% to 31\% for individual connection point pairs. In addition, values in the second washing and drying trial $d_2$ seem higher compared to previous and subsequent trials, possibly due to a measurement error. Further testing is necessary to investigate the effects of washing and drying more comprehensively, and this observed variability needs to be incorporated into the model design, so that gestures recognition is stable for any applications depending on it.

\subsection{Building Interactive Applications with Gesture-Recognizing Knitted Sensors}
\label{sec:application_potential}

\begin{figure}[h]
    \centering
    \includegraphics[width=\textwidth]{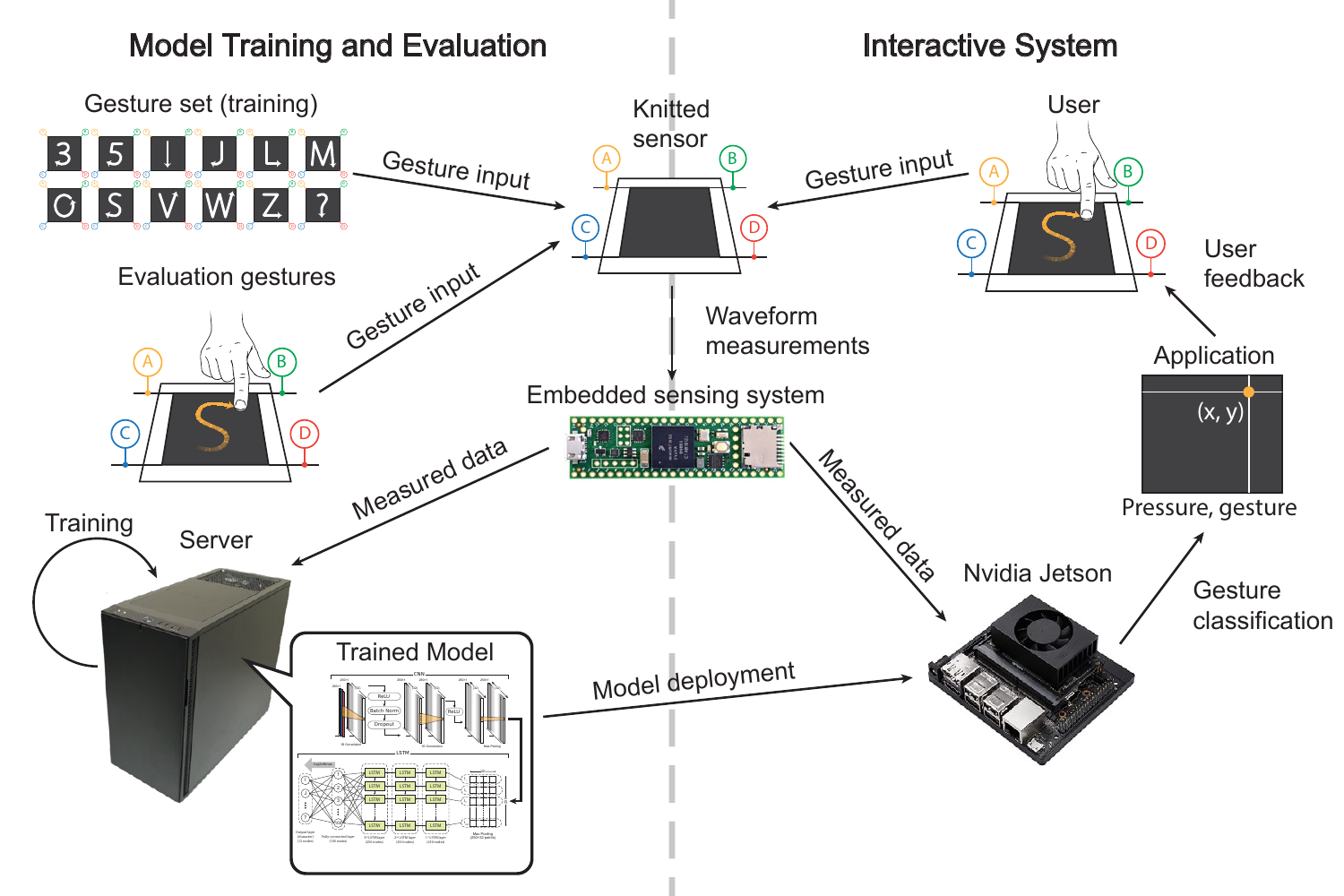}
    \caption{Processes and component interactions that describe the model creation and the working of an interactive gesture recognition system. Data collection, training and evaluation happen off-line and typically require more time and computing power. Once a model is trained to high accuracy, it can be deployed on lightweight hardware to recognize gestures in real time, supporting different interactive applications.}
    \label{fig:interactive_system}
    \Description{This figure illustrates two related process with their respective components: model training and evaluation on the left, and building an interactive system on the right. In the center, there are two components, a knitted sensor and embedded micro-controller, which are part of both processes. On the left, gesture sets for training and evaluation are depicted, gathered through user interaction with the knitted sensor. The sensor passes that information to the embedded micro-controller, and the data from there is used for training in a server computer. The server outputs a trained model, which is then deployed to the NVIDIA Jetson system-on-module, found on the right side of the figure, in the interactive system part. That process starts with the user performing a gesture on the knitted sensor, which is connected to the embedded micro-controller. The latter transmits the signal to the NVIDIA Jetson board, which, through its trained model, returns a gesture type. This gesture is then interpreted by an application, which gives a custom response to the user.}
\end{figure}

The system components described in Section~\ref{sec:system_design} create the foundation for building an interactive system that uses a knitted sensing area to accept gesture input and a machine learning model to determine the gesture performed by the user. In a real-time system, gestures can subsequently be interpreted by the specific application to trigger different events. This technology enables many kinds of applications, and additionally offers extensibility. Figure~\ref{fig:interactive_system} shows two related views of the system. On the left, it illustrates how data can be collected from users to train and evaluate a machine learning model. Gestures can be determined by the application needs, and the experiments and results in this work show that it is possible to recognize even relatively complex gestures, such as letters and numbers, with high accuracy. This allows great flexibility in the choice of gesture sets for training, even on a small-sized sensing area. Training happens offline on a server and after the cross-validation results achieve the required accuracy, the trained model is further evaluated, and then deployed on a smaller system, such as the NVIDIA Jetson computer. 

On the right side of Figure~\ref{fig:interactive_system}, the implementation of an interactive system based on this technology is illustrated. A user enters a gesture input on the knitted sensing area, connected to an embedded system for signal acquisition and pre-processing. This system would communicate in real time with the NVIDIA Jetson board hosting the trained gesture recognition model, capable of interpreting the signal as the gesture the user intended. The application relying on this technology would then respond to the user based on the meaning assigned to the particular gesture. 

This system configuration is extensible in the types of gestures recognized, since the model can be re-trained offline and re-deployed on existing hardware, allowing for large-scale production. Previous work~\cite{Vallett2019a,mcdonald2020knitted} has explored the potential of similarly-constructed sensors, and has introduced prototypes to illustrate their possible functionalities. The developments introduced in this work also hold promise for creating innovative applications. Some potential examples are broadly described below:

\begin{itemize}
    \item[-] \emph{Character Recognition:} The gesture examples used for recognition in this work are a subset of the character set of the English language. They were used to explore the feasibility of constructing a character recognition system. Having a system trained on the whole character set would allow written messages captured through fabric sensors to be transmitted. Since letters already have meaning embedded in them by convention, such an application would be intuitive, easy-to-use, and have great expressive power. 
    
    \item[-] \emph{Controllers:} Another category of applications that can be built using this sensor is that of controllers. This sensor allows the emulation of existing controller functionality, but with the added flexibility of fabric. For example, users can control their phone functionalities such as accepting or rejecting a phone call, changing the music, and more, through knitted interactive areas in their clothes, capable of recognizing gestures. Home automation is another potential application area, with such sensors integrated into furniture, pillows, or blankets, giving controllers a softer, more tactile-friendly quality. Gaming could benefit from application where knitted sensors are used as portable, foldable, and lightweight alternatives to hard-electronic controllers. Additionally, pressure sensitivity, an aspect of these sensors only explored in a limited way so far~\cite{Vallett2016a,Vallett2019a,mcdonald2020knitted}, could offer new interactive modalities for gaming and other applications.
    
    \item[-] \emph{Gestures in 3D Space:} Gestures do not need to be confined into a 2D plane. Knitted fabric can flex, fold, stretch, and move dynamically. Instead of only considering those fundamental aspects of fabric behaviour as qualities to design out, in order to maintain stability of touch or gesture representation, for certain applications we can also choose to design with these qualities at the center. For example, stretching could be given a specific meaning in a interactive system, and so can folding the fabric, twisting, or pinching it. An important aspect that needs to be considered while designing for such use cases however, is that capacitive sensing, the sensing strategy on which these sensors are designed, requires the presence of two conductors in proximity. In the typical cases, also explored in this work, the two conductors are the conductive yarn and human skin. 

    \item[-] \emph{Virtual and Augmented Reality (VR/AR) Applications:} Knitted sensors with gesture recognition technology integrated into them can be useful for VR/AR environments. It is easy to imagine objects that can be designed somewhat generically using knitted fabric with interactive gesture-sensing areas, with their functionality depending on the specific VR/AR application. This sensor's construction process allows for scalable design, resulting in interactive shapes capable of being built in different sizes. Therefore a whole environment could be composed of soft, knitted interactive objects, which take a different meaning and functionality, depending on the application and visuals overlayed on them. Additionally, such applications could be especially useful for kids, since they offer more safety than potential hard-electronic equivalents.
\end{itemize}

\section{Limitations and Future Work}  

The results from experiments with our sensing and recognition system demonstrate progress towards developing interactive textile gesture recognition systems. However, there are still several areas that warrant further investigation, some of which we discuss below. 

\subsection{User Studies for Increasing Model Capacity and Usability}
Gesture recognition accuracy still needs to be increased and more gestures need to be included. Even though each application might require its custom-defined gestures, in order to extend this system, it is necessary to ensure its feasibility and robustness with more subjects and more gesture types. Data collected using different fingers for performing gestures, different finger-placements on the sensing pad, different orientations of gesture trajectory, as well as subjects of different ages will need to be explored, due to possible changes in physiology which affect conductivity. Usability studies need to be performed to explore the potential of gesture-recognizing knitted sensors to be incorporated into end-users' everyday lives. Some potential applications were discussed in Section~\ref{sec:application_potential}. Future work will focus on further investigating possible uses of this technology, as well as building specific applications, testing their performance and usability in real-world scenarios, and getting user feedback about design aspects. 
\subsection{Resistance to Real-World Conditions}
The physical durability of the sensor should be more closely examined, since such sensors need to be able to withstand exposure to different weather and environmental conditions in everyday life. Experiments are needed to test the ability of the carbon-suffused nylon yarn to resists material aging and abrasion. Methods of surface enhancement and preservation, such as coating and lamination, should be investigated as potential solutions, and their effectiveness needs to be quantified. Additionally, the robustness of the trained models needs to be evaluated under conditions of possible distortion. Prior work~\cite{mcdonald2020knitted} investigated model stability for touch location identification under conditions of stretching and exposure to electromagnetic radiation for knitted sensors constructed using the same manufacturing process and carbon-coated yarn. This work investigated the effects of washing and drying on the sensor in Section~\ref{sec:washing_drying}. Further tests are necessary since those studies were limited in the number of people, as well as types of conditions. 

Another aspect to be considered is sensor performance while being worn, integrated into clothing, even though our limited study in Section~\ref{sec:wearing_sensor} demonstrated its robustness through encouraging results. Moving while wearing this sensor is expected to produce little to no distortion, and the study above included some light motion, as gestures were collected while the sensor was being worn. Despite this, more studies are necessary to explore the effect of intense physical activity, especially since such sensors are expected to find applications in athletics. Additionally, sweat, could possibly interfere with conductivity, since it contains electrolytes. If it seeps through the sensor, the overall resistance of the sensor will change, which is expected to affect the quality of acquired gesture data. Moreover, if a user is grounding himself or herself while touching a location on the sensor, his/her conductivity is increased, which will again affect the sensor response. False positives could be induced if a conductor came into contact with the sensing areas of the knitted component. Further studies are needed to fully investigate the extent of the effect of such conditions.

\subsection{Unexplored Interaction Modalities}
Other modalities of interaction enabled by the sensor introduced above, such as pressure sensitivity should be studied and implemented. For example, soft touch, which would cause a weak applied capacitance, could potentially pose a problem towards accurate gesture recognition. On the other hand, applications can be developed that use pressure differentiation as a feature. Additionally, current sensors do not recognize multi-touch, so only one gesture can be performed at a time on the knitted sensor. From experiments conducted with sensors built using a single conductive yarn and two connections, we know that if two locations are touched simultaneously, the generated signal appears to be coming from a point in between the two contact points~\cite{Vallett2019a}. In addition to these areas, we also plan to experiment with conductive yarns of different properties. The resistance of the yarn affects conductivity, and future models should also account for that quality.  

\section{Conclusion}

In this work, we create the foundations for building an interactive gesture recognition system using an easily-manufactured capacitive weft-knitted sensor, and a classification model that recognizes 12 English language characters as gestures. Additionally, we deploy the model on NVIDIA\textregistered~Jetson Xavier\TM~NX, a small, lightweight, and powerful embedded system-on-module.  

Previous work has shown several designs of similarly-produced sensors, which use weft knitting, an industry-standard manufacturing technique, for easy integration into clothes and other textile surfaces. Sensing areas of different shapes can be designed through this process, allowing for application-specific form factors. The sensor introduced in this work uses a single carbon-coated conductive yarn to create a continuous sensing area. That yarn is combined in the knitting process with other regular yarns to produce the final fabric substrate. Four electrodes are attached to the corners of the rectangular conductive area to measure voltage across it. Different gestures produce different voltage responses from the four connection points. 

In order to capture those identifying differences, we train and evaluate a CNN-LSTM neural network model with data from a total of 8 subjects. Leave-one-out cross-validation was used for training on 5 subjects, achieving a subject-independent accuracy of 78.6\%, and the data from the 3 other subjects was used for further evaluation of the model's generizability, during which, the model achieved a subject-independent accuracy of 89.8\%. Finally, we tested the response time of the trained model, to gesture data, after the model had been deployed on the Jetson system. The average response time was 30 ms, which is very encouraging to building a real-time application. These technical contributions help us advance toward feasible interactive embedded systems, capable of recognizing gestures on knitted senors.

Subsequently, we ran a user study to explore the model's accuracy response with gesture data collected while the sensor was being worn. Another experiment consisted of measuring the sensor's resistance across all connection points, after washing and drying it. These considerations, together with our discussion of the components and processes necessary to build real-time applications upon this technology help advance the capabilities of knitted textiles toward their use in real-world environments.

\begin{acks}
We thank the members of Drexel University’s Pennsylvania Fabric Discovery Center at the Center for Functional Fabrics for their invaluable digital knitting expertise and supervision of student volunteers. This research is supported in part by the Pennsylvania Fabric Discovery Center and the US Army Manufacturing Technology Program (US Army DEVCOM) under Agreement number W15QKN-16-3-0001.
\end{acks}

\bibliographystyle{ACM-Reference-Format}
\bibliography{textiles_ref,methods}

\appendix

\end{document}